\documentclass[prx,amsmath,amssymb, notitlepage, twocolumn,
nofootinbib,
superscriptaddress,
longbibliography
]{revtex4-1}
\usepackage{amsmath,amsfonts, amssymb, amsthm, dsfont}
\usepackage{yfonts}
\usepackage{bm}
\usepackage{mathrsfs}
\usepackage{graphicx}
\usepackage{verbatim}
\usepackage{tikz}
	\usetikzlibrary{calc}
	\usepackage{multirow}

\newcommand{\g}[1]{{}^{\ol{\mb{#1}}}}

\renewcommand{\vec}[1]{{\mathbf #1}}
\newcommand{\ket}[1]{|#1\rangle}

\renewcommand{\ol}[1]{\overline{#1}}
\newcommand{\comments}[1]{}
\renewcommand{\Ref}[1]{Ref. [\onlinecite{#1}]}

\newcommand{\mb}[1]{\mathbf{#1}}
\newcommand{\mc}[1]{\mathcal{#1}}

\newcommand{\beq}{\begin{eqnarray}}
\newcommand{\eeq}{\end{eqnarray}}
\newcommand{\ie}{{\it i.e., }}
\newcommand{\half}{\frac{1}{2}}
\newcommand{\trn}{{\rm trn}}
\newcommand{\hc}{{\rm h.c.}}
\newcommand{\eg}{{\it e.g.,\ }}
\newcommand{\Tr}{{\rm Tr}}

\usepackage{color}
\definecolor{darkblue}{rgb}{0.,0.,0.4}
\definecolor{darkred}{rgb}{0.5,0.,0.}
\usepackage{hyperref}

\newcommand{\spd}[1]{ (#1+1)\textrm{d}}
\def\TT{\mathsf{T}}

\def\U{\mathrm{U}(1)}
\def\UT{\mathrm{U}_\TT(1)}
\def\H{\mathcal{H}}
\def\Z{\mathbb{Z}}

\def\E{\mathsf{E}}
\def\M{\mathsf{M}}

\newcommand{\ga}[2]{ { {}^{\ol{\mb{#1}}}{#2} } }

\renewcommand{\vec}[1]{\bm{#1}}

\begin{document}

\title{Fractionalization and Anomalies in Symmetry-Enriched U(1) Gauge Theories}
\author{Shang-Qiang Ning}
\affiliation{Institute for Advanced Study, Tsinghua University, Beijing, 100084, China}
\affiliation{Department of Physics, The University of Hong Kong, Pokfulam Road, Hong Kong, China}
\author{Liujun Zou}
\affiliation{Department of Physics, Harvard University, Cambridge, MA 02138, USA}
\affiliation{Department of Physics, Massachusetts Institute of Technology, Cambridge, MA 02139, USA}
\affiliation{Perimeter Institute for Theoretical Physics, Waterloo, Ontario, Canada N2L 2Y5}
\author{Meng Cheng}
\affiliation{Department of Physics, Yale University, New Haven, CT 06511-8499, USA}


\begin{abstract}
   
   We classify symmetry fractionalization and anomalies in a \spd{3} U(1) gauge theory enriched by a global symmetry group $G$. We find that, in general, a symmetry-enrichment pattern is specified by 4 pieces of data: $\rho$, a map from $G$ to the duality symmetry group of this $\U$ gauge theory which physically encodes how the symmetry permutes the fractional excitations, $\nu\in\H^2_{\rho}[G, \UT]$, the symmetry actions on the electric charge, $p\in\H^1[G, \Z_\TT]$, indication of certain domain wall decoration with bosonic integer quantum Hall (BIQH) states, and a torsor $n$ over $\H^3_{\rho}[G, \Z]$,  the symmetry actions on the magnetic monopole. However, certain choices of $(\rho, \nu, p, n)$ are not physically realizable, \ie they are anomalous. We find that there are two levels of anomalies. The first level of anomalies obstruct the fractional excitations being deconfined, thus are referred to as the deconfinement anomaly. States with these anomalies can be realized on the boundary of a \spd{4} long-range entangled state. If a state does not suffer from a deconfinement anomaly, there can still be the second level of anomaly, the more familiar 't Hooft anomaly, which forbids certain types of symmetry fractionalization patterns to be implemented in an on-site fashion. States with these anomalies can be realized on the boundary of a \spd{4} short-range entangled state. We apply these results to some interesting physical examples.

\end{abstract}

\maketitle

\tableofcontents

\section{Introduction}

A three dimensional (\spd{3}) $\U$ quantum spin liquid (QSL) is an exotic gapless quantum phase. Due to the long-range entanglement inherent in this phase, it can be described by a compact \spd{3} $\U$ gauge theory at low energies {\footnote{In this paper we will use the terms ``$\U$ QSL" and ``$\U$ gauge theory" interchangeably.}}. It features emergent photons as the dominant low-energy excitations, but fractional excitations (\ie excitations with electric and/or magnetic charges) are still ineluctable in the system, even if they are gapped. This phase has been shown to be stabilized in a number of microscopic models \cite{Motrunich2002, Wen2003, Moessner2003, Hermele2004, Motrunich2004, Levin2006, Banerjee2008, Shannon2012}. Recently, the prospect of realizing $\U$ QSLs in the ``quantum spin ice'' phase of rare earth pyrochlores has stured much theoretical and experimental work \cite{Castelnovo2008, Morris2009, Fennell2009, Ross2009, Ross2011, Savary2011, Gingras2013}.

Microscopic realizations of a $\U$ QSL often enjoy certain global symmetries. In order to understand the physical properties of a $\U$ QSL, it is important to develop a systematic theory for the interplay between these global symmetries and its more intrinsic properties due to its long-range entanglement. As an example, the quantum spin ice has a time-reversal symmetry, and the monopoles are Kramers doublets under the time-reversal transformation, \ie the symmetry is realized projectively.

This understanding also provides useful information regarding the global phase diagram of a $\U$ QSL, especially its
proximate phases and the phase transitions between them.  For instance, condensation of electric or magnetic charges can drive the U(1) QSL to a short-range entangled phase, whose nature (\eg symmetry-breaking pattern) depends on the properties of the condensed charges~\cite{Motrunich2004}. Symmetry considerations are crucial in determining the properties of these proximate phases. 

\begin{figure}[h!]
	\centering
	\includegraphics[width=\linewidth]{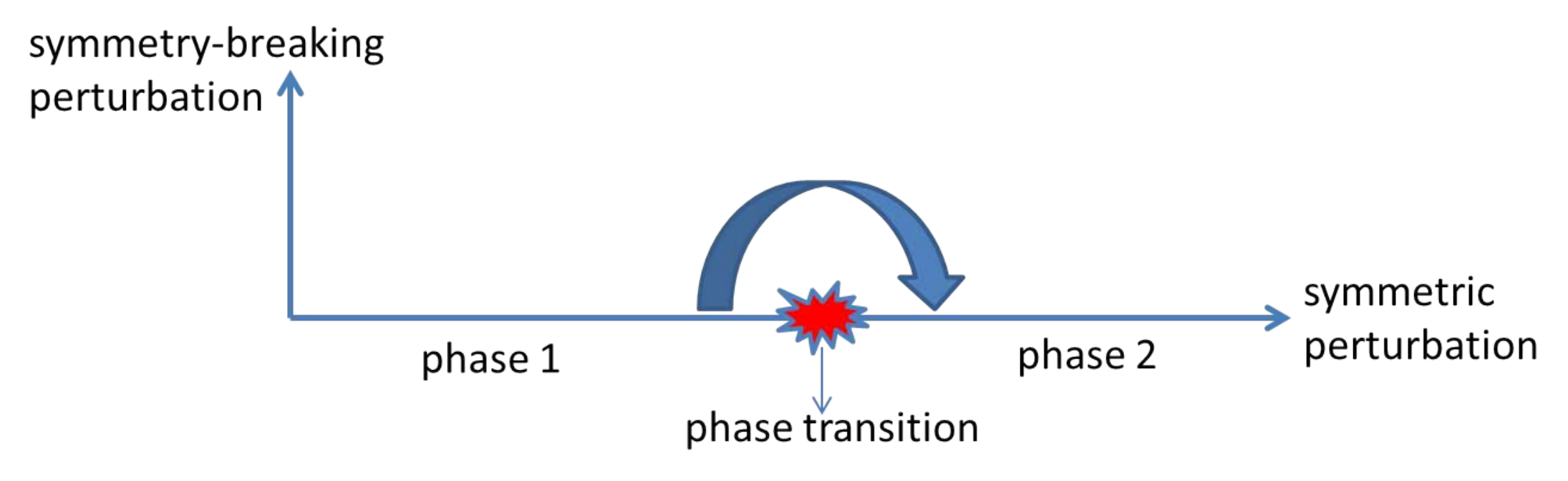}
	\caption{The notion of symmetry-protected distinction between two phases. These two phases can be smoothly connected if the system lacks certain symmetry, but they are necessarily separated by a phase transition in the presence of the symmetry.}
	\label{fig: symmetry-protected-distinction}
\end{figure}

A given set of global symmetries can have qualitatively distinct realizations in a $\U$ QSL, in the sense that $\U$ QSLs with different symmetry realizations can have symmetry-protected distinctions (see Fig. \ref{fig: symmetry-protected-distinction}). These different $\U$ QSLs are referred to as symmetry-enriched $\U$ QSLs under this symmetry.

Building on the preliminary work in Ref. \cite{Wang2013}, $\U$ gauge theories enriched by time reversal symmetry were first classified in Ref. \cite{Wang2015}. A systematic framework for the classification of generic symmetry-enriched $\U$ gauge theories was then proposed in Ref. \cite{Zou2017}, and this framework was applied to obtain the classifications of some rather nontrivial examples. In this framework, the bulk properties of a symmetry-enriched $\U$ gauge theory is characterized by statistics and the symmetry properties of the elementary electric charge and magnetic monopole of the theory, and its surface properties can be further enriched by weakly coupling it with a symmetry-protected topological (SPT) phase. In this paper, we will focus on the bulk properties of a symmetry-enriched $\U$ gauge theory.

To completely specify the symmetry properties of a U(1) QSL,  we need to know how symmetries act on the elementary electric charge and magnetic monopole, known as the symmetry fractionalization patterns. The symmetry actions on the elementary eletric charge and on the magnetic monopole are naively independent, but some of their combinations turn out to be anomalous, \ie a $\U$ gauge theory with certain symmetry fractionalization patterns cannot be realizable in any \spd{3} lattice spin system if the symmetry is implemented in an on-site manner, and it can only be realized as a boundary of a \spd{4} system. Ref. \cite{Zou2017} proposed a general physics-based method to detect such anomalies, and many nontrivial examples were demonstrated therein.

However, despite being general, systematic and physically intuitive, the method employed in Ref. \cite{Zou2017} can be sometimes sophisticated to implement. It is thus desirable to have a mathematical classification of anomalies, and a formula that indicates whether a symmetry fractionalization pattern is anomalous or not, and if it is anomalous, what kind of \spd{4} system can adopt this anomalous $\U$ gauge theory as its boundary. Furthermore, it is desirable if this anomaly formula can be formulated purely in terms of the physical symmetry quantum numbers of the elementary electric charge and magnetic monopole.

The main goal of this paper is to develop such a systematic understanding of anomalies in symmetry-enriched U(1) gauge theories.  As we will see, there are in fact two layers of anomalies: the first of them, the {\it deconfinement anomaly}, obstructs the deconfinment of the fractional excitations, rendering the notion of symmetry fractionalization ill-defined. When the first anomaly is absent, the second anomaly indicates whether the system has to live on the boundary of a \spd{4} nontrivial SPT phase. This is the more familiar 't Hooft anomaly.

The rest of the paper is organized as follows. In Sec. \ref{sec: review} we will give a brief review of the physics of a $\U$ gauge theory. In Sec. \ref{sec: anomaly formula}, after sketching its derivation, we will present a classification of symmetry-enriched U(1) gauge theories and the structure of their anomalies. This analysis is based on the conjecture that all anomaly-free symmetry-enriched $\U$ gauge theories can be viewed as partially gauged SPT phases. In this paper we will mostly consider $\U$ gauge theories with bosonic electric charges. We will then apply the anomaly formula to some interesting examples in Sec. \ref{sec: examples}. Some of these examples were discussed in Ref. \cite{Zou2017}, and our anomaly formula can reproduce the corresponding results and verify some conjectures made in Ref. \cite{Zou2017}. Besides these, we also discuss some other new intriguing examples. In particular, we discuss which $\U$ QSLs can be realized if SO(3) spin rotational symmetry and \spd{3} translation symmetry are preserved. Namely, we find symmetry-enriched $\U$ gauge theories that can satisfy the Lieb-Schultz-Mattis (LSM) constraint. We also discuss a symmetry-enriched $\U$ gauge theory that is related to the intrinsically interacting fermionic SPT phase found in Ref. \cite{ChengPRX2018}. Finally, we conclude in Sec. \ref{sec: discussion}. Various appendices contain some technical details.

\section{Review of $\U$ gauge theory} \label{sec: review}

\begin{figure}[h!]
	\centering
	\includegraphics[width=0.8\linewidth]{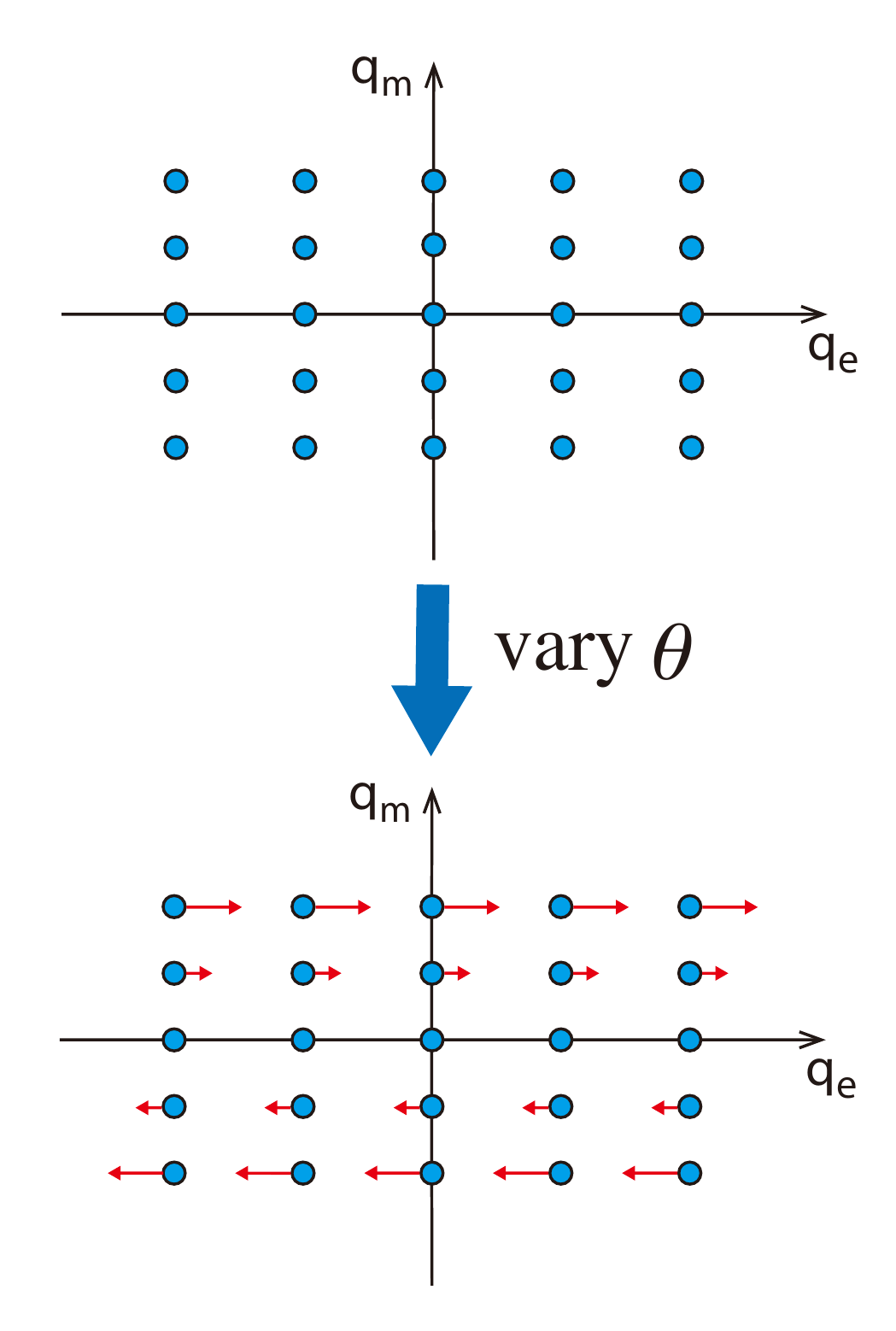}
	\caption{Upper: the possible values of the electric and magnetic charges of an excitation, $(q_e, q_m)$, form a charge-monopole lattice. This figure shows the charge-monopole lattice at $\theta=2\pi N$, where $N$ is an integer. Lower: When $\theta\neq 0$, the positions of the fractional excitations in the charge-monopole lattice are shifted due to the Witten effect. More precisely, the excitation with magnetic charge $q_m$ will get additional electric charge $\frac{\theta q_m}{2\pi}$. In the above figure, the lengths and directions of the red arrows indicate how the positions of the corresponding excitations change.}
	\label{fig: charge lattice and Witten effect}
\end{figure}

Generally a $\U$ gauge theory (with bosonic electric charge) is described by the following Lagrangian at low energies:
\begin{equation}
	\mathcal{L}=-\frac{1}{4e^2}f^{\mu\nu}f_{\mu\nu} + \frac{\theta}{32\pi^2} \varepsilon_{\mu\nu\lambda\rho}f^{\mu\nu}f^{\lambda \rho}
	\label{eqn:Lagrangian}
\end{equation}
Here $e$ is the gauge coupling strength and $\theta$ is the axion angle.  Notice that, in the absence of other symmetries, $\theta$ is $4\pi$-periodic if the charges are bosonic \cite{Vishwanath2012, Metlitski2013}. At low energies, the theory simply describes propagating photons. Above certain energy gap, there are fractional excitations carrying electric and magnetic charges. We denote the electric and magnetic charge of an excitation by $q_e$ and $q_m$, respectively. Due to the Dirac quantization condition \cite{Dirac1931}, the possible values of $q_e$ and $q_m$ form a charge-monopole lattice. Because of the $\theta$-term, an excitation acquires a ``polarization charge'' $\frac{\theta}{2\pi}q_m$ due to the Witten effect \cite{WITTEN1979} (see Fig. \ref{fig: charge lattice and Witten effect}). Therefore, the charge of a generic fractional excitation should be written as $q_e=n+\frac{\theta}{2\pi}q_m$, where $n$ is an integer counting the electric charge of this excitation at $\theta=0$. The self-statistics of a fractional excitation with electric and magnetic charges $(q_e, q_m)$ is given by $(-1)^{(q_e-\frac{\theta  q_m}{2\pi})q_m}$. This formula indicates that the statistics of the excitations is invariant when $\theta$ is changed by $4\pi$. 

In the absence of any orientation-reversing symmetries (time reversal and/or spatial reflection), $\theta$ can be tuned continuously. Without loss of generality, in this case we can always tune $\theta$ to be $0$ without encountering a phase transition. In the presence of an orientation-reversing symmetry, $\theta$ is quantized to be an integer multiple of $2\pi$. In all these cases, there is a charge-neutral monopole with a unit magnetic charge, \ie $q_m=1$. If $\theta=2\pi N$ with $N$ even (odd), the elementary charge-neutral monopole is bosonic (fermionic). We will denote by $\E$ the elementary electric charge with $(q_e, q_m)=(1, 0)$, denote by $\M$ the elementary charge-neutral monopole with $(q_e, q_m)=(0, 1)$. The charge-monopole lattice $\Z\times\Z$ is generated by $\E$ and $\M$, and we call bound states of certain numbers of $\E$ and $\M$ a dyon.

For $\theta=0\  {\rm(mod\ }4\pi)$, the $\U$ gauge theory has an emergent duality symmetry group of automorphisms, \ie permutations of fractional excitations that preserve all universal properties, such as exchange statistics. In defining automorphisms we ignore energetics such as gaps of the particles. Permutation of charges can be specified by its action on the two generators:
\begin{equation}
	\begin{pmatrix}
		q_e'\\
		q_m'
	\end{pmatrix}=
	\begin{pmatrix}
		a & b\\
		c & d
	\end{pmatrix}
	\begin{pmatrix}
		q_e\\
		q_m
	\end{pmatrix}.
	\label{}
\end{equation}
Clearly $a,b,c,d\in \Z$. In order to preserve the charge-monopole lattice, we must demand $ad-bc=\pm 1$. One can further show that only $ad-bc=1$ preserves the geometric Berry phase associated with braiding dyons, while $ad-bc=-1$ flips the Berry phase.  All integer $2\times 2$ matrices with unit determinant form the group SL$(2, \mathbb{Z})$, generated by
$\mathbf{S}$ and $\mathbf{T}$:
\begin{equation}
	\mathbf{S}=
	\begin{pmatrix}
		0 & -1\\
		1 & 0
	\end{pmatrix},
	\quad
	\mathbf{T}=
	\begin{pmatrix}
	    1 & 1 \\
	    0 & 1 \\
	\end{pmatrix}
	\label{}
\end{equation}
However, the $\mathbf{T}^n$ transformation changes the statistics of particles for odd $n$ (\eg a bosonic charge $(1,0)$ turns into a fermionic dyon $(n,1)$), so the group that preserves all Berry phases is actually generated by $\mathbf{S}$ and $\mathbf{T}^2$, and we will denote this group by $\mathcal{D}_+$. $\mathbf{T}^n$ with odd $n$ can only be realized in a U(1) gauge theories with fermionic charge, and we will not discuss them in this paper.

We can also consider the permutations reversing the sign of the Berry phase, which must correspond to orientation-reversing transformations. All these can be obtained from $\mathcal{D}_+$ by multiplying the following matrix
\begin{equation}
	\begin{pmatrix}
		1 & 0\\
		0 & -1
	\end{pmatrix}.
	\label{}
\end{equation}
We will denote all such permutations by $\mathcal{D}_-$.

Altogether, we have found the duality symmetry group $\mathcal{D}=\mathcal{D}_+ \oplus \mathcal{D}_-$.

Although we started from the relativistic Lagrangian Eq. \eqref{eqn:Lagrangian},  our  discussion  below  will  not  rely  on  Lorentz symmetry  in  essential  ways.   In  other  words,  we  consider more broadly quantum phases with emergent U(1) gauge symmetry with gapped electric and magnetic charges, which are not necessarily described by the relativistic Lagrangian.

\section{Symmetry fractionalization and anomalies in $\U$ gauge theory} \label{sec: anomaly formula}

Now we consider a $\U$ gauge theory realized in a microscopic model with a global symmetry group $G$. We will analyze how global symmetry transformations are realized in the low-energy theory. For clarity, let us assume that $G$ is internal, and we expect the results for spatial symmetries will be similar \cite{ThorngrenPRX2018, Zou2017a}. We will also consider the case where $G$ includes lattice translation symmetry in some occasions. Notice that $G$ may contain both unitary and anti-unitary transformations. To formally keep track of this, we define a $\Z_2$ grading $s: G\rightarrow \Z_2=\{1,-1\}$ on $G$ to indicate whether a group element $\mb{g}$ corresponds to a unitary ($s(\mb{g})=1$) or anti-unitary ($s(\mb{g})=-1$) transformation.

First of all, we consider how gauge-invariant operators transform under the symmetries. In the low-energy limit of a $\U$ gauge theory, all gauge-invariant local operators can be built up out of field strengths $\mb{E}$ and $\mb{B}$. They may transform nontrivially under a symmetry operation. For example, a charge conjugation symmetry takes $\mb{E}\rightarrow -\mb{E}$ and $\mb{B}\rightarrow -\mb{B}$. Equivalently, because $\mb{E}$ and $\mb{B}$ are sourced by electric and magnetic charges, we can also directly write down how the types of electric and magnetic charged excitations transform. In the example of charge conjugation, $\E\rightarrow\E^\dag$ and $\M\rightarrow\M^\dag$. Clearly such a transformation is an element in $\mathcal{D}$. Therefore, we have a group homomorphism $\rho$ from $G$ to $\mathcal{D}$ (preserving the grading $s$).

When $\rho$ is given, we still do not have a complete description of the symmetry action. The missing information is how symmetry acts locally on an individual fractional excitation, which will be referred to as symmetry fractionalization. A major goal of this work is to obtain a complete classification of both $\rho$ and symmetry fractionalization in physical $\U$ gauge theories. The basic principle is the following conjecture, first formulated in \Ref{Zou2017}:

\vspace{2mm}
\noindent\fbox{%
    \parbox{0.95\columnwidth}{%
\emph{All physical symmetry-enriched $\U$ gauge theories can be realized as partially gauged SPT phases}.
}
}
\vspace{2mm}

Let us elaborate on this statement. By physical, we mean that the $\U$ gauge theory can be realized in a 3D microscopic model with an on-site symmetry group $G$. The above conjecture allows us to only consider SPT phases whose symmetry group contains $\U$ as a normal subgroup, which after gauging becomes the $\U$ gauge symmetry. $G$ is the remaining global symmetry after gauging. We note that the above principle has also been applied to study symmetry-enriched SU$(N)$ gauge theories \cite{Guo2017, Wan2019}.

An immediate consequence of this conjecture is that one should be able to identify a certain dyonic excitation (and multiples of this dyon) as the matter of the SPT phase, coupled to a $\U$ gauge field. In the charge-monopole lattice, all the matters of this SPT phase should correspond to a line of lattice points passing through the origin. The global symmetry must fix this line in order for the gauging to make sense. Denote a dyon on this line by $(q_e, q_m)$, and suppose the symmetry transformation on the charge type is given by $\begin{pmatrix} a & b\\ c & d\end{pmatrix}$, then
\begin{equation}
	\begin{pmatrix}
		a & b\\
		c & d
	\end{pmatrix}
	\begin{pmatrix}
		q_e\\
		q_m
	\end{pmatrix}
	=k\begin{pmatrix}
		q_e\\
		q_m
	\end{pmatrix},
	\label{}
\end{equation}
where $k$ is a nonzero integer. To have a non-zero solution to this equation, we must have
\begin{equation}
		\det
	\begin{pmatrix}
		a-k & b\\
		c & d-k
	\end{pmatrix}=0.
	\label{}
\end{equation}
Together with $ad-bc=1$, we find $a+d=\frac{k^2+1}{k}$. Since $a+d\in \Z$, the only consistent choices are $k=\pm 1$, corresponding to $a+d=\pm 2$. In other words, such SL(2, $\Z$) matrices have trace $\pm 2$.  It is known that they are actually all conjugate to $\pm \mathbf{T}^n$ for $n\in \Z$. 
Because all such transformations have an infinite order except for $n=0$, when $G$ is a compact group (including finite groups) we only need to consider $n=0$, \ie the charge-conjugation subgroup. When $G$ contains an infinite-order element (\eg lattice translation), the element can act as the $\mb{T}$ transformations.
If the symmetry is realized with $n=0$, we can take any of the dyons as the SPT matter. {If the symmetry is realized with $n\neq 0$, we should take $(q_e, 0)$ as the SPT matter.}

We can also consider anti-unitary transformations, which have $ad-bc=-1$. Following a similar argument, we find that if there is a fixed line in the charge-monopole lattice, the trace must be $0$, \ie $a+d=0$. One can show that all such matrices are conjugate to either $\begin{pmatrix}1 & 0\\ 0 & -1\end{pmatrix}$ or $\begin{pmatrix}0 & 1\\ 1 & 0\end{pmatrix}$. The former case is just the usual convention that the electric (magnetic) fields are even (odd) under time reversal. In the later case, $(1,\pm 1)$ is the fermionic dyon identified as the SPT matter. This case corresponds to a $\U$ gauge theory with $\theta=\pi$. We will not consider this case further in this work.

To conclude this discussion, if we only consider compact symmetry groups, we may restrict the image of $\rho$ to the $\Z_2$ charge-conjugation subgroup of $\mathcal{D}$.

Next we analyze symmetry fractionalization.

\subsection{Symmetry fractionalization}
\label{sec:symfrac}

Based on the above discussion, in a $\U$ gauge theory a general compact symmetry group $G$ comes with a $\Z_2$-grading $\rho: G\rightarrow \Z_2=\{1,-1\}$. $\rho(\mb{g})=-1$ means $\mb{g}$ acts as charge conjugation:
\begin{equation}
	\mb{g}: \mb{E}\rightarrow -\mb{E},
	\quad
	\mb{B}\rightarrow -\mb{B}.
	\label{}
\end{equation}

Besides the charge-conjugation grading, there is also the $\Z_2$ grading $s$ to distinguish unitary and anti-unitary transformations. We will take the convention that the electric and magnetic fields transform as
\begin{equation}
	\mb{g}: \mb{E}\rightarrow \rho(\mb{g})\mb{E},
	\quad \mb{B}\rightarrow \rho(\mb{g})s(\mb{g})\mb{B}
	\label{eqn:fields_transform}
\end{equation}
So the transformation belongs to $\mathcal{D}_{s(\mb{g})}$. Equivalently, the charges transform as
\begin{equation}
	q_e\rightarrow {^{\mb{g}}q_e}=\rho(\mb{g})q_e, \quad q_m\rightarrow {^{\mb{g}}q_m}=\rho(\mb{g})s(\mb{g})q_m
	\label{}
\end{equation}

Once we specify how charges are permuted by symmetries, we examine how symmetry locally transforms an individual charge. Consider the action of the global symmetry operator $R_\mb{g}$ for $\mb{g}\in G$ on a physical state $\ket{\Psi}$ with multiple fractional excitations $a_1, a_2, \cdots, a_n$ which are spatially well-separated. The symmetry operator may transform the field lines induced by the charges, as given in Eq. \eqref{eqn:fields_transform}. In addition, $R_\mb{g}$ may also induce localized unitary transformations on each of the charges. We argue that
\begin{equation}
	R_\mb{g}\approx\prod_{j}U_\mb{g}^{(a_j)}\hat{\rho}_\mb{g}.
	\label{}
\end{equation}
Here we separate local unitary transformations $U_\mb{g}^{(a_j)}$ from the non-local transformation $\hat{\rho}_\mb{g}$ that acts globally on gauge theory. This equation should be understood as an (approximate) operator identity when operators localized in the neighborhood of charge excitations are concerned. 

Comparing the global symmetry action $R_\mb{g}R_\mb{h}$ and $R_\mb{gh}$ yields
\begin{equation}
\begin{split}
	 R_{\bf g} R_{\bf h}  &= R_{\bf g} \prod_{j=1}^{n} U^{(a_j)}_{\bf h} \hat{\rho}_{\bf h}
= R_{\bf g} \prod_{j=1}^{n} U^{(a_j)}_{\bf h} R^{-1}_{\bf g}R_{\bf g} \hat{\rho}_{\bf h} \notag \\
&= R_{\bf g} \prod_{j=1}^{n} U^{(a_j)}_{\bf h} R^{-1}_{\bf g} \prod_{k=1}^{n} U^{(a_k)}_{\bf g} \hat{\rho}_{\bf g} \hat{\rho}_{\bf h}    \\
& = \prod_{j=1}^{n} \,^{\bf g} U^{(a_j)}_{\bf h} U^{(a_j)}_{\bf g} \hat{\rho}_{\bf g} \hat{\rho}_{\bf h}  ,
\end{split}
\end{equation}
where $^{\bf g} U^{(a_j)}_{\bf h} = R_{\bf g} U^{(a_j)}_{\bf h} R^{-1}_{\bf g}$ has its nontrivial action localized within the vicinity of $a_j$, and we have used the fact {that $\hat{\rho}_\mb{g}\hat{\rho}_\mb{h}=\hat{\rho}_\mb{gh}$} and the fact that operators whose nontrivial actions are localized in different regions commute with each other.

Comparing this with $R_\mb{gh}=\prod_{j}U_\mb{gh}^{(a_j)}\hat{\rho}_\mb{gh}$, we must have
\begin{equation}
	{}^{\mb{g}}U_\mb{h}^{(a_j)} U_\mb{g}^{(a_j)}=\eta_{a_j}(\mb{g,h})U_\mb{gh}^{(a_j)},
	\label{}
\end{equation}
and
\begin{equation}
	\prod_{j=1}^n \eta_{a_j}(\mb{g,h})=1.
	\label{}
\end{equation}
In particular, $\eta_{-a}(\mb{g,h})=\eta_a(\mb{g,h})^{-1}$.

Now we consider the associativity:
\begin{equation}
	\begin{gathered}
	\begin{split}
		{}^\mb{gh}U_\mb{k}^{(a)}\, {}^{\mb{g}}U_\mb{h}^{(a)} U_\mb{g}^{(a)}&=\eta_a(\mb{g,h}){}^\mb{gh}U^{(a)}_\mb{k} U^{(a)}_\mb{gh}\\
	&=\eta_a(\mb{gh,k})\eta_a(\mb{g,h})U^{(a)}_\mb{ghk}.
	\end{split}\\
	\begin{split}
		{}^\mb{gh}U_\mb{k}^{(a)}\, {}^{\mb{g}}U_\mb{h}^{(a)} U_\mb{g}^{(a)} &= \eta_{{}^\mb{g}a}^{s(\mb{g})}(\mb{h,k}){}^\mb{g}U^{(a)}_\mb{hk} U^{(a)}_\mb{g}\\
		&=\eta_a(\mb{g,hk})\eta_{ {}^{\mb{g}}a }^{s(\mb{g})}(\mb{h,k})U^{(a)}_\mb{ghk},
	\end{split}
\end{gathered}
	\label{}
\end{equation}
So we have the associativity constraint
\begin{equation}
	\eta_a(\mb{g,hk})\eta_{^{\mb{g}}a}^{s(\mb{g})}(\mb{h,k})=\eta_a(\mb{gh,k})\eta_a(\mb{g,h}).
	\label{}
\end{equation}

There is some redundancy in $\eta_{a}(\mb{g,h})$ due to the freedom to redefine $U_\mb{g}^{(a)}$ by multiplying a phase $\zeta_a(\mb{g})$ to it. In order to not affect $R_\mb{g}$, they need to satisfy $\prod_j \zeta_{a_j}(\mb{g})=1$. This redefinition of local operators changes the phases $\eta_a(\mb{g,h})$ in the following way:
\begin{equation}
	\eta_a(\mb{g,h})\rightarrow \frac{\zeta_a(\mb{g})\zeta_{{}^{\mb{g}}a}^{s(\mb{g})}(\mb{h})}{\zeta_a(\mb{gh})}\eta_a(\mb{g,h}).
	\label{eqn:coboundary}
\end{equation}

Now let us specialize to $a=\E$ and $\M$. For $a=\E$, $\eta_{ {}^\mb{g}\E}(\mb{h,k})=\eta_\E^{\rho(\mb{g})}(\mb{h,k})$, therefore
\begin{equation}
	\eta_\E(\mb{g,hk})\eta_{\E}^{\rho(\mb{g})s(\mb{g})}(\mb{h,k})=\eta_\E(\mb{gh,k})\eta_\E(\mb{g,h}).
	\label{}
\end{equation}
$\eta_\E$ defines a $2$-cocycle in $Z^2_{\rho\cdot s}[G, \U]$, where the $\E$ subscript indicates that $G$ acts on $\U$ as identity/complex conjugation if $\rho\cdot s=1$ or $-1$. The redundancy in Eq. \eqref{eqn:coboundary} means that the equivalence classes of $\eta_\E$ are given by the second cohomology group $\H^2_{\rho\cdot s}[G, \U]$, which in the literature is also denoted as $\H^2_\rho[G, \mathrm{U}_\TT(1)]$, where the subscript $\TT$ indicates that the action of time reversal is given by $s$. A brief review of these mathematical concepts is provided in Appendix \ref{app:math}.

Similarly, we can show that $\eta_\M$ is classified by $\H^2_\rho[G, \U]$, where the $\rho$ subscript indicates that $G$ acts on $\U$ as identity/complex conjugation if $\rho=1$ or $-1$. So different symmetry fractionalization classes can be labeled by two $2$-cocycles $[\nu]\in \H^2_{\rho\cdot s}[G, \U]$ and $[\omega_\M]\in\H^2_\rho[G, \U]$. In Ref. \cite{Zou2017}, $[\nu]$ and $[\omega_\M]$ are dubbed the electric and magnetic projective representations of the symmetry group $G$, respectively.

Notice that in the absence of any orientation-reversing symmetry, the properties of the monopole can be changed by smoothly varying $\theta$.  To understand the effect of the $\theta$-term, let us start with $\theta=0$, where in our convention $\M$ is a boson with a certain projective quantum number $[\omega_\M]$. To get to the case with a nonzero $\theta$, we can imagine continuously tuning the value of $\theta$, so that the positions of the fractional excitations in the charge-monopole lattice are shifted due to the Witten effect (see Fig. \ref{fig: charge lattice and Witten effect}). To have a charge-neutral elementary monopole, we need to tune the value of $\theta$ to be an integral multiple of $2\pi$, say, $2\pi N$ with $N$ an integer. Then the projective quantum number of the charge-neutral elementary monopole with this value of $\theta$ is determined by the excitation with $(q_e, q_m)=(-N, 1)$ at $\theta=0$, which is $[\omega_\M\cdot\nu^{-N}]$ (this is well-defined since for an orientation-preserving symmetry both $\nu$ and $\omega_\M$ are classified by $\H^2_\rho[G, \U]$).
In particular, when $\theta$ is varied by $4\pi$, the statistics of the monopole is invariant, but its symmetry fractionalization pattern gets shifted by $[\nu^{-2}]$. So in this case $[\omega_\M]$ is well-defined only up to $[\nu^2]$.

In the presence of an orientation-reserving symmetry, $\theta$ is quantized to be a multiple of $2\pi$. In this case, we can still define the symmetry fractionalization class of charge-neutral monopoles (not just up to $[\nu^2]$).

This discussion can be generalized to other duality transformations, e.g. the $\mb{S}$ transformation, as long as they are compactible with the symmetry action on charge types. Two symmetry fractionalization classes related by these duality transformations should correspond to the same symmetry-enriched phase. For example, if $G$ is unitary and acts as identity or charge conjugation, the $\mb{S}$ transformation commutes with the symmetry action, so in this case the electric and magnetic projective representations can be interchanged without changing the phase.

Therefore, following general considerations, we have found that a symmetric $\U$ gauge theory is equipped with 4 pieces of data: symmetries permuting charge types, given by $\rho$, and projective symmetry transformations, parametrized by $\nu\in\H^2_{\rho\cdot s}[G, \U]$, the value of $\theta$, and $\omega_\M\in\H^2_\rho[G, \U]$. However, it is not clear that every $(\rho, [\nu], \theta, [\omega_\M])$ can be realized physically in \spd{3} as a partially gauged SPT phase. In fact, it is sometimes even problematic to discuss the classification of projective quantum numbers if action of $\rho$ contains e.g. $T^2$ elements. Below we address this issue.

\subsection{Ungauging} \label{sec: gauged SPT}

We would like to construct a U(1) gauge theory from gauging a bosonic SPT phase. First, we must identify the matter particles, or ``gauge charges''. Suppose that the matter is generated by a particular dyon $(q_e, q_m)$. Without loss of generality, we may assume $\text{gcd}(q_e, q_m)=1$. We will further assume that $(-1)^{q_eq_m}=1$, so the dyon is bosonic. Then we perform a duality transformation so that this dyon becomes the $\E$ charge $(1,0)$. More explicitly, the duality transformation takes the following form:
\begin{equation}
	U=\begin{pmatrix}
		x & y\\
		-q_m & q_e
	\end{pmatrix},
	\label{}
\end{equation}
with $xq_e+yq_m=1$. It is straightforward to check that $U\begin{pmatrix} q_e\\ q_m\end{pmatrix}=\begin{pmatrix} 1\\ 0\end{pmatrix}$. The ``monopole'' $(0,1)$ is actually the image of $(-y, x)$ under this duality transformation. Notice that since $x,y$ are not uniquely determined, there are infinitely many choices of the ``monopole'', which are related to each other via $T$ transformations. From now on, we will assume that such a duality transformation has been done, so that the matter is generated by the bosonic $\E$ charge, and there is a charge-neutral elementary monopole $\M$. In the absence of any orientation-reversing symmetry, $\M$ will always be taken as a boson, because this can be achieved by smoothly tuning $\theta$ to be 0.  On the other hand, in the presence of a orientation-reversing symmetry, the value of $\theta$ cannot be smoothly varied and the statistics of $\M$ is a robust universal feature of this symmetry-enriched phase.

We now determine the structure of the symmetry group of the matter.  In the $\U$ gauge theory, the $\E$ charge can transform projectively under $G$, with a factor set $\nu$ that specifies the corresponding projective representation. Correspondingly, in the SPT phase the fundamental charge-$1$ boson carries the same projective representation of $G$. Mathematically, it means that the actual symmetry group $\mathcal{G}$ of the SPT phase is an extension of $G$ by $\U$ (while the symmetry group of the $\U$ gauge theory is of course just $G$). For notational convenience we use $\U$ and its isomorphic group $\mathbb{R}/2\pi\Z$ interchangeably, \ie $e^{i\theta}\in\U$ is identified as $\theta\in[0, 2\pi)$.  Let us now define $\mathcal{G}$. Denote the unitary transformation associated with $\mb{g}\in G$ by $R_\mb{g}$, and let $R_\theta=e^{iQ\theta}$ be a $\U$ rotation. We have the following relation:
\begin{equation}
	R_\mb{g}R_\theta R_\mb{g}^{-1}=R_{s(\mb{g})\rho(\mb{g})\theta}.
	\label{eqn:group-relation-1}
\end{equation}
Because charged bosons transform as projective representations of $G$, we have
\begin{equation}
	R_\mb{g}R_\mb{h}=e^{i\nu(\mb{g,h})Q} R_\mb{gh}.
	\label{eqn:group-relation-2}
\end{equation}
 These two relations Eqs. \eqref{eqn:group-relation-1} and \eqref{eqn:group-relation-2} completely determine the group structure of $\mc{G}$.  In the following it will be more convenient to use additive notations for group multiplication, and label elements of $\mc{G}$ as $a_\mb{g}$ where $a\in \mathbb{R}/2\pi\Z$ and $\mb{g}\in G$. The multiplication in $\mc{G}$ is then given by
\begin{equation}
	a_\mb{g}\times b_\mb{h}=[a+{}^{\mb{g}}b+\nu(\mb{g,h})]_\mb{gh}.
	\label{}
\end{equation}
with $ {}^{\mb{g}}x=\rho(\mb{g})s(\mb{g})x$.

It is now well-understood that the classification of bosonic SPT phases in $d=1, 2$ and $3$ spatial dimensions is given by group cohomology $\H^{d+2}[\mathcal{G}, \Z_\TT]$~\cite{chen2013}, plus additional ``beyond cohomology'' phases when anti-unitary symmetries are present in 3D given by $\H^1[\mathcal{G}, \Z_\TT]$~\cite{Vishwanath2012, Xiong_2018, Gaiotto2017}. For compact (or finite) $\mathcal{G}$, we have $\H^{d+2}[\mathcal{G}, \Z_\TT]\simeq \H^{d+1}[\mathcal{G}, \UT]$. The ``beyond cohomology'' part is not relevant for our purpose, because $\H^1[\mathcal{G}, \Z_\TT]\simeq \H^1_s[G, \Z]$ and describes SPT phases protected by $G$ alone.  Below we present an explicit description of $\H^4[\mathcal{G}, \UT]$.

\subsubsection{Projective quantum numbers of monopoles}\label{subsec: monopole rep}

Before discussing the general classification, we first explain how the symmetry properties of the magnetic monopole is encoded in this formalism.

Let us start from the simplest case where $G$ is unitary and $\rho=\mathds{1}, [\nu]=[0]$. In this case, $\mathcal{G}=\U\times G$. The K\"unneth formula then implies
\begin{equation}
	\begin{split}
	\H^4[\U\times G, \U]
	&=\H^4[G, \U]\times\H^3[G, \Z]\\
	&=\H^4[G, \U]\times\H^2[G, \U].
	\end{split}
	\label{}
\end{equation}
The last equality assumes a compact/finite $G$.
Physically, $\H^4[G, \U]$ describes SPT phases protected by $G$ alone and thus is not of interest. The other factor, $\H^2[G, \U]$, describes projective representations of $G$ and it is very natural to identify it with the fractionalization class $[\omega_\M]$ of magnetic monopoles. In fact, we can find the following explicit parametrization of $4$-cocycle in $\H^4[\mathcal{G}, \U]$:
\begin{equation}
	\omega(a_\mb{g}, b_\mb{h}, c_\mb{k}, d_\mb{l})=\mu(\mb{g,h,k,l})e^{ia n(\mb{g,h,k})}.
	\label{eq: 4-cocycle simple}
\end{equation}
Here $\mu$ must be a $4$-cocycle of $G$, and
$n$ is a $3$-cocycle in $\H^3[G, \Z]\simeq\H^2[G, \U]$.

We will show that the $3$-cocycle $n$ indeed encode the symmetry fractionalization pattern on the monopole, \ie it is equivalent to $\omega_\M$. To do so, let us first turn on the $\U$ gauge field. In the group-cohomology models with a unitary symmetry group, a $4$-cocycle in fact determines the space-time partition function on a general $4$-manifold~\cite{Kapustin2014}, equipped with background gauge fields~\footnote{When the symmetry group is finite, such a partition function can be rigorously represented as a finite state sum on a triangulated manifold. Although in our case the symmetry group contains $\U$ and more work is needed to rigorously write down the partition function, we will dispense mathematical rigor for now and proceed formally.}. Since the SPT phase is gapped and we are only interested in the topological part of the response theory, we can assume that the gauge fields are flat. Denote the partition function of the SPT phase on a closed space-time manifold $M$ equipped with the background $\U$ gauge field, represented by a $\mathbb{R}$-valued 1-cochain $A$, and $G$ gauge field $g$, by $\mathcal{Z}(M; A, g)=\exp\left(iS_\text{top}[M; A,g]\right)$. The expression of $S_\text{top}[M; A,g]$ is determined by Eq. (\ref{eq: 4-cocycle simple}), and it is given explicitly in Eq. (\ref{eqn:Stop1}). If $A$ is promoted to be a dynamical gauge field, then the partition function in the presence of a background $G$ gauge field is
\beq
\mc{Z}(M; g)=\int\mc{D}A\,e^{iS_\text{top}[M; A,g]+iS[M;A]},
\eeq
where $S[M; A]$ includes both the Maxwell term and the $\theta$-term $S_\theta[A]$, which contains no coupling between $A$ and $g$. All coupling between $A$ and $g$ is in the topological term:
\begin{equation}
  S_\text{top}[M;A,g]=\int_M A\cup n.
  \label{eqn:Stop1}
\end{equation}
Here $n(g)$ is the $\Z$-valued 3-cocycle on $M$ which is the pull-back of $n\in \H^3[G, \Z]$ by the map $g: M\rightarrow BG$ corresponding to the gauge field $g$.  This is essentially equivalent to Eq. \eqref{eq: 4-cocycle simple}. Notice in writing the above action, we have dropped terms that only depend on $g$ (and $M$). These terms physically describe attaching a $G$-SPT to the $\U$ gauge theory, and they will not be considered in this paper.

Now using the correspondence between $\H^3[G, \Z]$ and $\H^2[G, \mathbb{R}/\Z]$, we write $n=\frac{1}{2\pi}\beta\omega$ with $\omega\in Z^2[G, \mathbb{R}/\Z]$ where $\beta$ is the Bockstein homomorphism. Using integration by parts we find
\beq
S_\text{top}[M; A,g]=\frac{1}{2\pi}\int_MF\cup\omega(g).
\eeq
Here $F=\delta A$ is the field strength.
Formally this action is analogous to the well-known $F\wedge F$ topological theta term, and it will potentially give the monopole nontrivial projective quantum number under $G$.

To fully unearth the physical consequence of $S_\text{top}[M;A,g]$,  we put the theory on $M=S^2\times M_2$, with $S^2$ containing only spatial components and $M_2$ a general space-time $2$-manifold, and put a $2\pi$ $\U$ flux through $S^2$ (\ie $S^2$ encloses a unit monopole). We then take a limit where the linear size of $M_2$ is much greater than that of the $S^2$. Now this partition function describes the quantum amplitude of a process in which a monopole moves in the reduced spacetime $M_2$. This quantum amplitude receives contributions from both $S_\text{top}[M; A, g]$ and the $\theta$-term. The contribution from the $\theta$-term is analyzed above in Sec. \ref{sec:symfrac}: the $\theta$-term can change the projective quantum number of the monopole by $-\frac{\theta}{2\pi}[\nu]$.  The contribution from $S_{\rm top}$ becomes $\frac{1}{2\pi}\int_MF\cup\omega=\int_{M_2}\omega$. This means that the worldline of the monopole is further associated with an additional contribution to the quantum amplitude, $\int_{M_2}\omega$, which is precisely the $M_2$ partition function of a \spd{1} $G$-SPT state whose boundary realizes the projective representation specified by the factor set $[\omega]$. That is to say, the magnetic flux line is further decorated with this \spd{1} $G$-SPT state, and its end point, the magnetic monopole, gets one more piece of contribution to its projective representation of $G$, which is specified by the factor set $[\omega]$. So when both the $\theta$-term and the $S_{\rm coupling}$ are taken into account, the projective quantum number of the charge-neutral monopole is given by $[\omega]-N[\nu]$ for $\theta=2\pi N$. In the present case, $[\nu]=0$ and $[\omega_M]$ is just identified as $[\omega]$.

\subsection{Structure of $\H^4[\mathcal{G}, \UT]$}
\label{sec:H4}
Now we explain the main result of this work, the structure of the cohomology group $\H^4[\mathcal{G}, \UT]$. Details of the proofs of our statements can be found in the Appendix \ref{app:4cocycle}. 

Recall that $\mathcal{G}$ is an extension of $G$ by $\U$, with a $2$-cocycle $[\nu]$.
A cohomology class in $\H^4[\mathcal{G}, \UT]$ is specified by three layers of data:
\begin{enumerate}
    
    \item A 1-cocycle $[p]$ from $\H^1_s[G, \Z]$, where the coefficient $\Z$ is in fact $\H^3[\U, \U]$. In other words, $p\in \Z$ satisfies $p(\mb{g})+s(\mb{g})p(\mb{h})=p(\mb{gh})$. Importantly, $[p]$ and $[\nu]$ need to satisfy an obstruction-vanishing condition: Define 
    \begin{equation}
      \gamma(\mb{g,h,k})=2s(\mb{gh})[\nu(\mb{g,h})]_{2\pi}p(\mb{k}).  
    \end{equation}
    Here $[x]_{2\pi}$ represents the fractional part of $x$ with respect to $2\pi$, \ie $[x]_{2\pi}=x\ \text{mod\ } 2\pi$ and $[x]_{2\pi}\in[0, 2\pi)$. One can easily show that $\gamma$ is a 3-cocycle in $Z^3_\rho[G, \mathbb{R}/2\pi\Z]$.
    There must exist $n\in C^3_\rho[G, \Z]$ such that $\delta_\rho n=\Gamma$, where $\Gamma=\frac{1}{2\pi}\beta_\rho \gamma\in Z_\rho^4[G, \Z]$. Namely, $\gamma$ needs to be a trivial cocycle in $\H^3_\rho[G, \mathbb{R}/2\pi\Z]$ for this obtruction to vanish. We will call $[\gamma]$ the $\H^3$ {\it deconfinement obstruction} (or symmetry localization obstruction) class, for reasons that will become clear later. We remark that this obstruction class is purely determined by $\rho$ (how the symmetry permutes fractional excitations), $[\nu]$ (the symmetry actions on the electric charge $\E$), and $[p]$, whose meaning will be explained below. In contrast, the symmetry actions on the magnetic monopoles are not in charge of this obstruction, as will be clear later.
    
    \item When the deconfinement obstruction vanishes, we can solve $\delta_\rho n=\Gamma$, and different solutions of $n$ are parametrized by a torsor over $\H^3_\rho[G, \Z]$. When $\nu=0$, the obstruction class is canonically zero, and we have shown that $n$ describes projective representation carried by magnetic monopoles in Sec. \ref{subsec: monopole rep}. Based on the mathematical structure, we conjecture that the same interpretation holds more generally, namely, the $\H^3_\rho[G, \Z]$ torsor classifies symmetry fractionalization on monopole excitations.
    
    \item Finally, an obstruction $5$-cocycle $[\mathcal{O}]\in \H^5_s[G, \U]$ must vanish. Otherwise, the $\U$ gauge theory that would arise from gauging this SPT phase must be realized on the boundary of a (4+1)d SPT phase defined by $[\mathcal{O}]$. When $[\mathcal{O}]$ is trivial, we may modify the 4-cocycle by an element from $\H^4_s[G, \U]$, corresponding to stacking a G-SPT phase. Notice that this does not necessarily lead to a new symmetry-enriched $\U$ gauge theory \cite{Wang2015, Zou2017}. The full expression for $[\mathcal{O}]$ is rather complicated and is given in Eq. \eqref{eqn:h5full} of 
    Appendix \ref{app:4cocycle}. Below we will consider in detail specific cases where the formula simplifies significantly.
    
\end{enumerate}

To better understand the classification, we consider a few simplified cases.  

\vspace{1mm}

\noindent \textbf{Case 1:} If $G$ is unitary and compact (finite), then $\H^1[G, \Z]=\Z_1$, so we can set $p=0$, which implies that the obstruction class $\Gamma$ vanishes identically. In this case, the 4-cocycle has the following simple representation:
\begin{align}
    \omega(a_\mb{g},b_\mb{h},c_\mb{k},d_\mb{l})=\omega(\mb{g,h,k,l})e^{ian(\mb{h}, \mb{k}, \mb{l})}.
\end{align}
As before, $\omega(\mb{g}, \mb{h}, \mb{k}, \mb{l})$ is a 4-cocycle in $\H^4_s[G, \U]$, and $n$ can be taken as a 3-cocycle in $\H^3_\rho[G, \Z]$.

As before, the 3-cocycle $n$ encodes the information of the symmetry fractionalization class on the monopole. Using $[n]$ and $[\nu]\in\H^2_{\rho\cdot s}[G, \U]$, which characterizes the symmetry fractionalization class on the charge, we have the following expression for the obstruction $5$-cocycle:
\beq \label{eq: obstruction 5-cocycle simple main text}
\mc{O}(\mb{g,h,k,l,m})=e^{-is(\mb{gh}){^{\overline{\mb{gh}}}\nu}(\mb{g}, \mb{h}) n(\mb{k}, \mb{l}, \mb{m})}.
\eeq
We claim that a $\U$ gauge theory with symmetry fractionalization pattern given by $\rho$, $[\nu]\in\H^2_{\rho\cdot s}[G, \mathbb{R}/2\pi\Z]$ and $[n]\in\H^3_\rho[G, \Z]$  is realizable if and only if $[\mc{O}]$ belongs to the trivial class in $\H^5_s[G, \U]$. If $[\mc{O}]$ belongs to a nontrivial class in $\H^5_s[G, \U]$, then this $\U$ gauge theory is anomalous, and can only be realized on the boundary of a \spd{4} G-SPT characterized by the 5-cocycle $[\mc{O}]$. We will provide further arguments for this statement in Sec. \ref{subsec: informal}.
\vspace{1mm}

\noindent \textbf{Case 2:}
For a general finite/compact group $G$ that contains anti-unitary elements, we have $\H^1_s[G, \Z]=\Z_2$. It is not difficult to show that $p$ must take the following form:
\begin{equation}
    p(\mb{g})=k\cdot \frac{1-s(\mb{g})}{2}
    \label{eqn:p}
\end{equation}
with $k$ an integer. Even (odd) $k$ represents the trivial (nontrivial) class of $\H^1_s[G, \Z]$. Let us further assume $[\nu]=0$ for simplicity,
and consider the following 4-cocycle 
\begin{equation}\label{eqn:case 2 4-cocycle}
    \omega(a_\mb{g}, b_\mb{h},c_\mb{k}, d_\mb{l})=
    \omega_{\mb{k,l}}^{s(\mb{gh})}(\g{gh}a,\g{h}b)\omega_\mb{l}^{s(\mb{ghk})}(\g{ghk}a,\g{hk}b,\g{k}c).
\end{equation}
Here $\omega_\mb{g}(a,b,c)$ is given by
\begin{equation}
    \omega_\mb{g}(a,b,c)=e^{\frac{i}{2\pi}p(\mb{g})a([b]_{2\pi}+[c]_{2\pi}-[b+c]_{2\pi})}.
\end{equation}
We note that $[\omega_\mb{g}]\in \H^3[\U, \U]$  describes a bosonic integer quantum Hall (BIQH) state of Hall conductance $\sigma_{xy}=2p(\mb{g})$.
The expression for $\omega_{\mb{k,l}}(a,b)$ can be found in Appendix \ref{app:4cocycle}. Eq. \eqref{eqn:case 2 4-cocycle} is not the most general form of 4-cocycle in this case, but the following explanation holds more generally.

We claim that this $4$-cocycle {with $p(\mb{g})$ given by Eq. \eqref{eqn:p}} corresponds to $\theta=2\pi k$. 
To see it, consider the slant product of $\omega(a_\mb{g}, b_\mb{h}, c_\mb{k}, d_\mb{l})$ over $\mb{g}$ (see Appendix \ref{app:math} for a brief introduction of slant products):
  \begin{equation}
    (i_{\mb{g}}\omega)(a,b,c)=\omega_\mb{g}(a,b,c). 
    \label{eqn:slantong}
  \end{equation}
It is well-known that the slant product corresponds to dimensional reduction of the system onto a domain wall~\cite{WangLevinPRB2015}.
Eq. \eqref{eqn:slantong} means that the quantum state on a domain wall labeled by $\mb{g}$ can be described by the data $\omega_\mb{g}(a,b,c)$. From Eq. (\ref{eqn:p}), we see that when $s(\mb{g})=1$, $p(\mb{g})=0$, the slant product gives $1$ and this domain wall is a  trivial state. On the other hand, when  $s(\mb{g})=-1$, $p(\mb{g})=k$, and the domain wall is a bosonic integer quantum hall state with $\sigma_{xy}=2k$ (in units of $e^2/h$) \footnote{Notice when a coboundary transformation on $\omega(a_{\mb{g}}, b_{\mb{h}}, c_{\mb{k}}, d_{\mb{l}})$ is performed, $(i_{\mb{g}}\omega)(a,b,c)$ can change by $e^{\frac{i}{\pi}a([b]_{2\pi}+[c]_{2\pi}-[b+c]_{2\pi})}$, the 3-cocycle that describes a BIQH state with $\sigma_{xy}=4$. This precisely reflects the fact that the states with $\theta=4\pi$ and with $\theta=0$ are in the same phase in the absence of any symmetry other than $U$ and time reversal.}. This exactly matches the properties of a state with $\theta=2\pi k$ \cite{Senthil2012, Vishwanath2012, XuPRB2013}. So, intuitively, the 4-cocycle can be interpreted as decorating (2+1)d BIQH states onto time-reversal domain walls, classified by $\H^3[\U, \U]=\Z$. We should emphasize that this {relation between $p(\mb{g})$ and $\theta$} only holds for anti-unitary symmetry $\mb{g}$. If $\mb{g}$ is unitary, in general there is no such relation between $[p]$ and the $\theta$-term. Also, notice that Eq. \eqref{eqn:p} only holds for compact/finite symmetry groups, and it does not hold for symmetries like lattice translations, which will be discussed next.
\vspace{1mm}

\noindent \textbf{Case 3:} For lattice translation symmetry along the $z$ direction, $\Z$, we have $\H^1[\Z, \Z]=\Z$, and $p$ can take any integral value. We call $\mb{z}$ the element that translates the system by one lattice spacing along the $z$ direction. Again consider the 4-cocycle given by Eq. \eqref{eqn:case 2 4-cocycle}, and now the meaning of the slant product Eq. \eqref{eqn:slantong} is that on each plane perpendicular to the $z$ direction, we have a BIQH with $\sigma_{xy}=2p(\mb{z})$~\cite{ChengPRX2016}. Therefore, $p$ is a more general concept than the $\theta$ value, and it indicates certain domain wall decoration with BIQH states. In this case of translation symmetry, in the corresponding $\U$ gauge theory the action of translation is $\mathbf{T}^2$ \cite{Williamson_arxiv}. 
To see this, consider a magnetic monopole in the system. When the monopole is translated by one unit along $z$, say from below $z=0$ to right above $z=0$, the magnetic flux through the plane $z=0$ changes by $-2\pi$. The quantum Hall response then creates charge-$(-2p)$ on the $z=0$ plane. As a result, the under $T_z$ the monopole transforms as
\begin{equation}
	\M\rightarrow \M {\E^{\dag}}^{2p}.
\end{equation}
That is, this translation transformation is in fact $\mathbf{T}^{2p(\mb{z})}$.
We notice that the theory obtained by gauging the U(1) symmetry of an infinite stack of BIQH states has unconventional low-energy dynamics, as the photon dispersion is no longer linear~\cite{Williamson_arxiv}. Therefore it can not be described using the relativistic Lagrangian Eq. \eqref{eqn:Lagrangian}~\footnote{$\mb{T}^2$ transformation is not an explicit symmetry of the relativistic theory \eqref{eqn:Lagrangian}.}

\subsection{$\H^3$ obstruction class}
Having discussed the meaning of $n$ and $p$, we now further elaborate on the $\H^3$ obstruction class. 

First of all, if $G$ is a finite group or a compact Lie group, the general form of $p$ is given by Eq (\ref{eqn:p}). When $G$ is unitary, $s(\mb{g})=1$ and $p(\mb{g})=0$ for all $\mb{g}$, so the $\H^3$ obstruction class vanishes.

Now consider a general $s(\mb{g})$, \ie the symmetry group may contain anti-unitary elements. It turns out that even for a general $s(\mb{g})$, the $\H^3$ obstruction also vanishes identically. To see it, define $u(\mb{g,h})=-s(\mb{gh})[\nu(\mb{g,h})]_{2\pi}$. It is straightforward to show that $\delta_\rho u=\gamma \text{ mod }2\pi\mathbb{Z}$, so $\Gamma=\frac{1}{2\pi}\beta_\rho\gamma$ vanishes. 

Let us demonstrate why this is the case physically by considering an example with $G=H \times \Z_2^\mathsf{T}$, where $H$ is unitary and finite. We denote the group element of $H$ as $h$, and $\Z_2^\mathsf{T}=\{1,\mathsf{T}\}$. Let us also suppose  that $\nu$ entirely comes from $H$. We choose $p$ as in Eq. (\ref{eqn:p}) with $k>0$. We will also set $\rho\equiv \mathds{1}$ in this example. Notice so far we have only specified the data responsible for the $\H^3$ obstruction class, and our discussion is independent of the possible presence of the $\H^5$ obstruction class.

A $\mathcal{G}$-SPT phase can always be obtained by first breaking the $\U$ symmetry and making the system a superfluid, and then proliferating the vortex lines of this superfluid. In order for such a gapped state to exist, a vortex line to be proliferated must be fully gapped without any degeneracy or gapless modes.  

Since $p(\mathsf{T})=k$, a BIQH state is decorated onto a time-reversal domain wall. Suppose we thread a $2\pi$ flux through the domain wall. Due to the $\sigma_{xy}=2k$ quantum Hall response, the flux threading creates a charge-$2k$ excitation, which carries a $G$ projective representation labeled by $2k\nu$. 
In other words, on a $2\pi$ flux line a $\mathsf{T}$ domain wall binds a ``zero mode" protected by the $G$ symmetry (in this example, $H$) when $2k\nu$ is nontrivial.  Naively, this poses an obstruction to proliferating vortex lines to yield a gapped symmetric state, as the proliferation seems to break the $G$ symmetry.

However, we are allowed to decorate the vortex lines to be proliferated with gapped 1D states. In this example, we can just decorate the vortex lines with a 1D $H$-SPT phase with a factor set $k\nu$. Due the time-reversal symmetry domain wall, the two sides on the vortex lines have 1D SPT states labeled by $k\nu$ and $-k\nu$, with a $-2k\nu$ projective representation sitting on the domain wall and neutralizing the projective representation arising from the Hall response. Now everything is gapped, and it is possible to proliferate the vortex lines to get a symmetric gapped state, if the $\H^5$ obstruction class further vanishes.
In fact, as long as no symmetry acts as $\mb{T}$-transformations, the corresponding \spd{3} symmetry-enriched $\U$ gauge theory can at most suffer from an $\H^5$ obstruction, and it can always be realized on the boundary of a \spd{4} invertible state \cite{Zou2017} (see Appendix C therein), which can be constructed via a generalization of the layer construction in Ref. \cite{Wang2013}.
    
Now we give an example where the $\H^3$ obstruction class is actually nontrivial. We choose the symmetry group to be $G=H\times \Z$. Notice this is not a compact/connected Lie/finite group. Denote the generator of $\Z$ by $\mb{z}$. Consider an example with $p(\mb{z})=k$.  To see whether the $\H^3$ obstruction class is nontrivial in this case, we compute the slant product $i_z \gamma|_{H}=2k\nu$. As long as $2k\nu$ is nontrivial, the $\H^3$ obstruction class is nontrivial.

To have a concrete example, suppose $H=\mathrm{PSU}(N)$ with $N>2$ (or its finite subgroup $\Z_N\times\Z_N$).  If we take $\nu$ to be the fundamental representation of $\mathrm{SU}(N)$ (the generating element in $\H^2[\mathrm{PSU}(N),\U]=\Z_N$), then in order for the $\H^3$ obstruction class to vanish, we need
\begin{equation}
k=
    \begin{cases}
    N/2, & N\text{ is even}\\
    N, & N\text{ is odd}
    \end{cases}.
    \label{eqn:kconstraint}
\end{equation}

We can interpret the $\Z$ as lattice translation. As explained in the previous section, such a $\mathcal{G}$-SPT phase can be viewed as a stack of 2D BIQH phases with Hall conductance $\sigma_{xy}=2k$.  However, since the matter boson carries the fundamental representation of SU$(N)$, the Hall conductance is constrained to be a multiple of $N$ ($2N$) when $N$ is even (odd) (see Appendix \ref{app: Hall} for derivation). This is exactly the condition that the $\H^3$ obstruction class $\gamma$ vanishes. 

If $k$ takes any other integer value, then the $\H^3$ obstruction class is nontrivial, which means the state with those other values of $k$ are not valid $\mc{G}$-SPTs. Let us understand what is wrong with those states. Suppose such a state could be realized, then we can gauge the $\U$ symmetry to obtain a $\U$ gauge theory. After gauging, whenever a $2\pi$ magnetic flux line goes through a plane of such a BIQH state, charge-$2k$ will be left on the plane due to the nonzero Hall conductance. This charge-$2k$ object carries projective representation $2k\nu$ of the PSU$(N)$ symmetry, thus resulting in symmetry-protected degeneracy (gapless modes) on this magnetic flux line. Note that in this case we cannot cancel the degeneracy by attaching  \spd{1} PSU$(N)$ SPT state on the $2\pi$ magnetic flux line. In a $\U$ gauge theory, the $2\pi$ magnetic flux lines need to be ``condensed" for the monopoles to be deconfined. However, the presence of the gapless modes makes these flux lines visible, and, as a result, the monopoles cannot be viewed as deconfined excitations, which contradicts our assumption that this state can be gauged to yield a $\U$ gauge theory. For this reason, we refer to the $\H^3$ obstruction as the {\it deconfinement obstruction}. Because now the monopoles are not deconfined excitations, it does not make sense to talk about localizing symmetry actions on them, and such an obstruction can also be called a symmetry localization obstruction.

So what sort of \spd{4} bulk can support such an \spd{3} SPT phase on the boundary? To answer this, let us first ask what sort of \spd{3} bulk can support on its boundary a BIQH with $\sigma_{xy}$ violating the constraint given in Eq. \eqref{eqn:kconstraint}.  In Appendix \ref{app: Hall}, we show that a \spd{3} bulk with the following $\theta$-term in the response can produce the desired response on its \spd{2} surface:
\begin{equation}
    S[M_{4}; A]=\frac{2\pi\sigma_{xy}}{8\pi^2}\int_{M_4} F\cup F.
\end{equation}
Strictly speaking, the U(1) gauge field $A$ needs to satisfy additional conditions to reflect the fact that charges carry projective representations of PSU($N$), see Appendix \ref{app: Hall} and Sec. \ref{subsec: informal} for for details.
This type of \spd{2} states are referred to as anomalous invertible states \cite{Wang2018}. Namely, this invertible state can only exist on the boundary of a higher-dimensional \emph{trivial} bulk. If we try to gauge the U(1) symmetry in the anomalous invertible state, the dynamical gauge field resulting from gauging also has to be extended into the bulk. 
 
Now we come back to the 3D stack of the 2D anomalous BIQH states, and ask on the boundary of what kind of \spd{4} bulk this \spd{3} stack can be realized. Apparently, the \spd{4} bulk that supports the
anomalous \spd{3} invertible phase must also contain topological terms. Suppose the \spd{4} space-time manifold is $M_{5}$. Formally, if we introduce a $\mathbb{Z}$ gauge field $z\in H^1[M_{5}, \Z]$, the bulk response is given by
\begin{equation}
    S[M_{5};A,z]=\frac{2\pi\sigma_{xy}}{8\pi^2}\int_{M_5} F\cup F\cup z.
\end{equation}

If we place the \spd{4} theory on $S^2\times M_3$, and let $\int_{S^2}F=2\pi$, the partition function then yields the following theory living on a ``flux surface" (or the worldsheet of a ``monopole" loop in four spatial dimensions):
\begin{equation}
    \sigma_{xy} \int_{M_3} F\cup z.
\end{equation}
As we explain below, because electric charges carry projective representations, we need to identify $F=\frac{2\pi}{N}w(B)$, where $B$ is the background PSU$(N)$ bundle, and $w(B)\in H^2[M_3, \Z_N]$ is the characteristic class that describes the obstruction of lifting a PSU$(N)$ bundle to SU($N$) bundle. So the action is essentially $\frac{2\pi\sigma_{xy}}{N}\int w(B)\cup z$, which describes the Lieb-Schultz-Mattis anomaly of a \spd{1} PSU$(N)$-symmetric spin chain, where each site transforms as the projective representation labeled by $\sigma_{xy}$~\cite{Yao2018}, as expected from the physical argument presented earlier as the flux surface terminates on a flux line on the \spd{3} boundary. Indeed, these PSU$(N)$-symmetric spin chains live on the boundary of \spd{2} SPTs classified by $\H^3[G, \U]$, which is also the classification of the $H^3$ anomalies here. 

From this example, we see that a natural way to resolve a non-vanishing $\H^3$ deconfinement obstruction is to require that both the background $G$ gauge field and the dynamical $\U$ gauge field be extended to the higher-dimensional bulk, and is therefore quite different from the usual 't Hooft anomaly. 
This is similar to the symmetry-localization obstruction found in (2+1)d symmetry-enriched topological phases~\cite{SET,  Tarantino_SET, FidkowskiPRB2017, BarkeshliPRB2018}.

\subsection{'t Hooft anomaly formula}\label{subsec: informal}

Before finishing this section, we will sketch an informal derivation of the 't Hooft anomaly formula in the special case where $\rho(\mb{g})=s(\mb{g})=1$ for all $\mb{g}\in G$, which also explains the physical meaning of the object given by Eq. \eqref{eq: obstruction 5-cocycle simple main text}. We will limit ourselves to the case $p=0$.

Suggested by the explicit parametrization, we postulate that the topological response theory of the to-be-gauged SPT takes a form similar to Eq. \eqref{eqn:Stop1}:
\begin{equation}
	S[M_4; A, g]=\int_{M_4} A\cup n.
	\label{eq: 4D response}
\end{equation}
While we still use the notation $A,g$ to represent the $\mathcal{G}$ background gauge field, we must keep in mind that $\mathcal{G}$ is generally not a direct product of $\U$ and $G$. In particular it means that one has to modify the flat connection condition to
\begin{equation}
	\delta A = \nu(g).
	\label{}
\end{equation}
Here $\nu$ is the pull-back of the group $2$-cocycle $\nu\in \H^2[G, \mathbb{R}/2\pi\Z]$ to the $G$ bundle.

The response has to be gauge-invariant.  Under a $G$ gauge transformation, $\nu(g)$ is shifted by $\delta f_1$ where $f_1$ is a 1-cochain, and $n$ is shifted by $\delta f_2$. We do not need to know the specific forms of $f_1$ and $f_2$. In order to preserve the flatness of the gauge field, $A$ must be shifted to $A+f_1$. Therefore, the topological response theory changes by
\begin{equation}
	\begin{split}
	\int_{M_4} &(f_1\cup n + A\cup \delta f_2 + f_1\cup \delta f_2)\\
	&=\int_{M_4} (f_1\cup n + \nu\cup f_2 + f_1\cup \delta f_2).
	\end{split}
	\label{eqn:variation}
\end{equation}
Here we used $\delta (A\cup f_2) = \delta A \cup f_2-A\cup \delta f_2$. Thus the theory is not gauge-invariant. But the variation is now seen to only depend on the $G$ gauge field. This suggests that we fix the problem by including a 5D bulk $M_5$ whose boundary is $M_4$, with the following action:
\begin{equation}
	S[M_5; g]=-\int_{M_5}\nu \cup n.
	\label{eq: 5D response}
\end{equation}
Here $g$ is an extension of the $G$ gauge field to $M_5$. Notice this \spd{4} response theory is essentially Eq. \eqref{eq: obstruction 5-cocycle simple main text}.

Let us check that the variation of $S_5$ under a $G$ gauge transformation does give Eq. \eqref{eqn:variation}:
\begin{equation}
	\begin{split}
	&-\int_{M_5} [(\nu + \delta f_1)\cup (n+\delta f_2)-\nu\cup n] \\
	=&-\int_{M_5}(\delta f_1\cup n + \nu \cup \delta f_2 + \delta f_1\cup \delta f_2)\\
	=&-\int_{M_5}\delta (f_1\cup n + f_1\cup \delta f_2 + \nu \cup f_2)\\
	=&-\int_{M_4} (f_1\cup n + f_1\cup \delta f_2 + \nu \cup f_2)
	\end{split}
	\label{}
\end{equation}
Thus this term exactly cancels Eq. \eqref{eqn:variation}.

Therefore the whole theory (5D bulk and 4D boundary) is gauge-invariant. Since the 5D bulk response only depends on the $G$ gauge field, it describes a $G$-SPT phase. This result means that the U(1) gauge theory obtained by gauging the $\mc{G}$-SPT described by the 4D action Eq. \eqref{eq: 4D response} can live on the boundary of a 5D $G$-SPT phase described by Eq. \eqref{eq: 5D response}.

\section{Applications} \label{sec: examples}

In this section we will apply the anomaly formula to various examples. In all these examples, the deconfinement obstruction class always vanishes.

\subsection{$G=\Z_2$} \label{subsec: Z2}

Let us first consider $\U$ gauge theories enriched by a unitary $\Z_2$ symmetry. The extension of $\Z_2$ by $\U$ is given by
\begin{equation}
	\H^2_\rho[\Z_2, \U]=
	\begin{cases}
		\Z_1 & \rho=\mathds{1}\\
		\Z_2 & \rho=-\mathds{1}
	\end{cases}.
	\label{}
\end{equation}
Physically, $\rho=-\mathds{1}$ means that the $\Z_2$ symmetry acts as a charge conjugation, and $\rho=\mathds{1}$ means that it does not act as a charge conjugation.

For the case with $\rho=\mathds{1}$, because $\H^2_\rho[\Z_2, \U]=\Z_1$, there is no nontrivial symmetry fractionalization pattern, and there is only one possible $\U$ gauge theory with no fractionalization on $\E$ or $\M$. This state is denoted by $\E_b\M_b$ in Ref. \cite{Zou2017}. 

Below we will study the $\rho=-\mathds{1}$ case. A representative 2-cocycle is:
\begin{equation}
	\eta(\mb{g,g})=(-1)^\lambda.
	\label{}
\end{equation}
with $\lambda=0,1$.  In the notions of Ref. \cite{Zou2017}, the cases with $(\lambda_\E, \lambda_\M)=(0, 0)$, $(\lambda_\E, \lambda_\M)=(1, 0)$ or $(\lambda_\E, \lambda_\M)=(0, 1)$, and $(\lambda_\E, \lambda_\M)=(1, 1)$ are denoted as $(\E_b\M_b)_-$, $(\E_{bZ}\M_b)_-$ and $(\E_{bZ}\M_{bZ})_-$, respectively.

Let us compute the obstruction $5$-cocycle. From $\eta\equiv e^{i\omega}$ we find the only non-zero component of $n$:
\begin{equation}
	n(\mb{g,g,g})=\omega(\mb{g,g})- {}^\mb{g}\omega(\mb{g,g})=2\omega(\mb{g,g})=\lambda.
	\label{}
\end{equation}
Then using the formula Eq. \eqref{eq: obstruction 5-cocycle simple main text} we have
\begin{equation}
	\mathscr{O}(\mb{g,g,g,g,g})=(-1)^{\lambda_\E \lambda_\M}.
	\label{}
\end{equation}
This is a nontrivial 5-cocycle if and only if $\lambda_\E=\lambda_\M=1$. So $(\E_b\M_b)_-$ and $(\E_{bZ}\M_{b})_-$ are anomaly-free, while $(\E_{bZ}\M_{bZ})_-$ is anomalous and must be realized on the boundary of a \spd{4} group-cohomology $\Z_2$ SPT phase. Indeed, there is a \spd{4} group-cohomology $\Z_2$ SPT phase, and our result implies that $(\E_{bZ}\M_{bZ})_-$ can be its boundary state. These results agree with Ref. \cite{Zou2017}.

We notice that in (4+1)d there is a ``beyond-cohomology'' $\Z_2$ SPT phase \cite{Wen2014}. The boundary of this phase is characterized by a mixed $\Z_2$-gravity anomaly. We now argue that the boundary can not be a $\Z_2$-symmetry-enriched U(1) gauge theory. First of all, we may restrict to U(1) gauge theories with both electric and magnetic charges bosonic (the other possibility, the so-called all-fermion electrodynamics, has global gravitational anomaly). Since the symmetry is $\Z_2$, it either acts trivially or as charge conjugation on charge types, as these are the only order-two elements in the duality group. 
Under these conditions, all possible $\Z_2$-symmetry-enriched U(1) gauge theories have been exhausted here, therefore we conclude that this ``beyond-cohomology'' SPT phase cannot have a U(1) gauge theory as symmetry-preserving boundary termination.

\subsection{$G=\mathrm{SO}(3)$}
\label{sec:so3}
Let us now consider an example of an anomalous U(1) QSL with SO(3) spin rotational symmetry. Ref. \cite{Zou2017} shows that the state $\E_{b\half}\M_{b\half}$, where both $\E$ and $\M$ are bosons that carry spin-1/2, is anomalous.

Now we apply our obstruction formula to re-derive this result. It suffices to show that this state is still anomalous when the SO(3) symmetry is broken down to its $\Z_2\times\Z_2$ subgroup, consisting of three $\pi$ rotations around $x,y$ and $z$ axes. This is the minimal subgroup of SO(3) where the spin-$1/2$ projective representation still makes sense, since $\H^2[\Z_2\times\Z_2, \U]=\Z_2$. In the anomalous theory, both $\E$ and $\M$ carry the nontrivial projective representation of $\Z_2\times\Z_2$. Ref. \cite{Zou2017} suggested that this state is still anomalous, and we indeed find that the obstruction class is nontrivial, thus verifying this statement. The details will be postponed to Sec. \ref{sec:fermionspt}.

\subsection{$G=\Z_2\times\Z_2^{\mathsf{T}}$}

Next we consider the symmetry group $G=\Z_2\times\Z_2^\mathsf{T}$. This symmetry is relevant for experimental QSL candidates made of non-Kramers quantum spins. Ref. \cite{Zou2017} found 75 symmetry fractionalization patterns for $\U$ gauge theories with this symmetry, where 38 of them are anomaly-free and the other 37 are anomalous. We will apply our anomaly formula to rederive the anomalies of the 37 anomalous states, and we will also confirm a conjecture made in Ref. \cite{Zou2017} about the anomaly classes.

Let us denote $G=\{1, Z, \TT, Z\TT\}$ where $Z$ is the generator of the $\Z_2$ subgroup and $T$ the generator of the $\Z_2^{\mathsf{T}}$ subgroup. They satisfy $Z^2=\TT^2=1, Z\TT=\TT Z$. The homomorphism $\rho$ is determined by $\rho(Z)$. We can then systematically classify fractionalization classes (see Appendix \ref{app: Z2 x Z2T} for details).

Let us consider how to distinguish cohomology classes in $\H^5[G, \U]$. Applying K\"unneth formula, we find
\begin{equation}
	\begin{split}
		\H^5[\Z_2\times\Z_2^\mathsf{T}, \U]=&
	\H^5[\Z_2, \H^0[\Z_2^\mathsf{T}, \U]]\oplus\\
	&\H^3[\Z_2, \H^2[\Z_2^\mathsf{T}, \U]]\oplus\\
	&\H^1[\Z_2, \H^4[\Z_2^\mathsf{T}, \U]]
	\end{split}
		\label{}
\end{equation}
Given a $5$-cocycle $\mathcal{O}$, we can decompose the cohomology class in the following way:
\begin{equation}
	[\mathcal{O}]=r_1[\mathcal{O}_1] + r_2[\mathcal{O}_2] + r_3 [\mathcal{O}_3],
	\label{}
\end{equation}
where $r_{1,2,3}\in \{0,1\}$, and $\mathcal{O}_1$ is the generating class of $\H^5[\Z_2, \H^0[\Z_2^\mathsf{T}, \U]]=\H^5[\Z_2, \U_1]$, $\mathcal{O}_2$ for $\H^3[\Z_2, \H^2[\Z_2^\mathsf{T}, \U]]$, $\mathcal{O}_3$ for $\H^1[\Z_2, \H^4[\Z_2^\mathsf{T}, \U]]$. $\mathcal{O}_1$ corresponds to \spd{4} SPT phases protected by $\Z_2$ alone, which is precisely the state whose boundary can be $(\E_{bZ}\M_{bZ})_-$ (see Sec. \ref{subsec: Z2}). Below we will focus on the remaining $\Z_2^2$ part.

We now discuss how to determine $r_i, i=2,3$, from $\mathcal{O}$.  We consider $r_3$ first, which turns out to be simpler to define. We use a cohomology operation called slant product, which for each group element $\mb{g}$ defines a group homomorphism $i_\mb{g}:\H^{d}[G, \U]\rightarrow \H^{d-1}[G, \U]$ (see Appendix \ref{sec:slant} for a review). We define $(-1)^{r_3}=(i_Z \mathcal{O})(\mathsf{T,T,T,T})$.

To find $r_2$, we need a generalization of slant product, 2-slant product, which are defined now for multiple group elements, see Appendix \ref{sec:slant}. We define 
\begin{equation}
	(-1)^{r_2}=(i_{\mathsf{T,T}}\mathcal{O})(Z,Z,Z).
	\label{}
\end{equation}
Using the definition of 2-slant product in Appendix \ref{sec:slant}, one can check that both $r_2$ and $r_3$ are invariants for the cohomology class (\ie invariant under coboundary transformations).

\begin{table}[h]
\begin{tabular}{c|c|c|c|c}
\hline\hline
${\footnotesize\rho(Z)}$ & ${\footnotesize \nu(\mb{g,h})}$  & ${\footnotesize \omega_\M(\mb{g,h})}$ & {\footnotesize Anomaly class}  &  {\footnotesize Notation in Ref. \cite{Zou2017}} \\
\hline
\multirow{2}{*}{$1$}& $(-)^{g_2h_2}$ & $(-)^{g_1h_2}$  & $(0,0,1)$  &$\E_{bTT'}\M_{b-}$  \\
    & $(-)^{g_1h_1}$ & $(-)^{g_1h_2}$ & $(0,1,0)$  &$\E_{bT'}\M_{b-}$ \\
    \hline
 \multirow{4}{*}{$-1$}&   $(-)^{g_1h_1}$ &   $(-)^{g_1h_1}$ & $(1,0,0)$ & $(\E_{bZ}\M_{bT'Z})_-$  \\
 & $(-)^{g_1h_1}$  &  $(-)^{g_2h_2}$ &  $(0,1,0)$ &  $(\E_{bZ}\M_{bT'})_-$ \\
 &$(-)^{g_2h_2}$  &  $(-)^{g_1h_1}$ &  $(0,1,0)$ &  $(\E_{bT}\M_{bT'Z})_-$ \\
 & $(-)^{g_2h_2}$  &  $(-)^{g_2h_2}$ &  $(0,0,1)$ &  $(\E_{bT}\M_{bT'})_-$ \\
 \hline \hline
\end{tabular}
\caption{Anomaly classes for a couple of $\U$ gauge theories with $\Z_2\times\Z_2^{\mathsf{T}}$ symmetry.}
\label{tab:z2z2t}
\end{table}

We compute the obstruction classes when both $\nu$ and $\omega_\M$ are nontrivial (when either of them is trivial the obstruction class vanishes automatically). The result is tabulated in Table \ref{tab:z2z2t}. 

Ref. \cite{Zou2017} indeed found that all the 6 states we consider here are anomalous. In fact, after exhausting all possible symmetry fractionalization patterns of this symmetry, Ref. \cite{Zou2017} found in total 37 anomalous $\Z_2\times \Z_2^\mathsf{T}$ symmetric $\U$ gauge theories. Furthermore, the arguments therein (see Sec. VII C of Ref. \cite{Zou2017}) imply that, to show the anomalies of all these 37 states, it actually suffices to show that $(\E_{bZ}\M_{bZ})_-$, $(\E_{bT}\M_{bT'})_-$ and $\E_{bTT'}\M_{b-}$ are anomalous, which we have shown here. Therefore, we have reproduced the results in Ref. \cite{Zou2017} on anomalous $\Z_2\times\Z_2^\mathsf{T}$ symmetric $\U$ gauge theories. 

Ref. \cite{Zou2017} also conjectured a classification of the anomaly classes of these 37 anomalous states, within each class the anomaly of the states are the same. Our results also confirm this conjecture. More precisely, there are 6 anomaly classes (see Ref. \cite{Zou2017} for the properties of these states):
\begin{itemize}

\item[1.] $(\E_{bZ}\M_{bZ})_-$, $(\E_{bTZ}\M_{bT'Z})_-$, $(\E_{fT}\M_{bZ})_-$, $(\E_{bZ}\M_{fT'})_-$, $(\E_{fT}\M_{bT'Z})_-$, $(\E_{bTZ}\M_{fT'})_-$, $(\E_{fT}\M_{fT'})_{\theta-Z}$.

\item[2.] $(\E_{bTZ}\M_{bZ})_-$, $(\E_f\M_{bZ})_-$, $(\E_{bTZ}\M_f)_-$.

\item[3.] $(\E_{bZ}\M_{bT'Z})_-$, $(\E_{bZ}\M_f)_-$, $(\E_f\M_{bT'Z})_-$.

\item[4.] $(\E_{bZ}\M_{bT'})_-$, $(\E_{fTZ}\M_{bT'})_-$, $(\E_{fTZ}\M_{bT'Z})_-$, $(\E_{bT}\M_{bT'Z})_-$, $(\E_{bT}\M_{fZ})_-$, $(\E_{bZ}\M_{fZ})_-$, $\E_{bT}\M_{f-}$, $\E_{fT'}\M_{b-}$, $\E_{bT}\M_{b-}$.

\item[5.] $(\E_{bT}\M_{bZ})_-$, $(\E_{bT}\M_{fT'Z})_-$, $(\E_{bTZ}\M_{fT'Z})_-$, $(\E_{bTZ}\M_{bT'})_-$, $(\E_{fZ}\M_{bT'})_-$, $(\E_{fZ}\M_{bZ})_-$, $\E_{bT'}\M_{b-}$, $\E_{bT'}\M_{f-}$, $\E_{fT}\M_{b-}$.

\item[6.] $(\E_{bT}\M_{bT'})_-$, $(\E_{f}\M_{bT'})_-$, $(\E_{bT}\M_f)_-$, $\E_{bTT'}\M_{b-}$, $\E_{bTT'}\M_{f-}$, $\E_f\M_{b-}$.

\end{itemize}

\subsection{Lieb-Schultz-Mattis-Hastings-Oshikawa anomaly}
\label{sec:LSManomaly}

We now apply our results to systems in which Lieb-Schulz-Mattis-Hastings-Oshikawa (LSMHO) type theorems~\cite{LSM, OshikawaLSM, HastingsLSM} hold. For concreteness, consider a translation-invariant lattice with spin-$1/2$ per unit cell, whose symmetry group is SO(3)$\times \Z^3$. The LSMHO theorem states that such a system does not allow a non-degenerate ground state preserving all symmetries on a torus. Such a constraint can be understood as the manifestation of a particular 't Hooft anomaly, if we view this lattice system as the boundary of a \spd{4} crystalline SPT ``bulk'' that consists of a stack of Haldane chains in the 4th dimension~\cite{ChengPRX2016, JianPRB2018, HuangPRB2017, MetlitskiPRB2018}. We will refer to this anomaly as the LSM anomaly. Our goal is to understand the implication of such an anomaly in a U(1) gauge theory.

Let us first explicitly write down the ``bulk'' theory for the LSM anomaly. While the protecting symmetry involves lattice translations, we will nevertheless treat them formally as an internal symmetry and imagine coupling the bulk to gauge fields of the translation symmetries $\Z^3$, denoted by $x,y,z$ for translations in the three orthogonal directions. We also turn on a background SO(3) gauge field $B$. The bulk response theory takes the following form~\cite{MetlitskiPRB2018}:
\begin{equation}
	S_\text{LSM}[M_5; B,x,y,z]=\pi\int_{M_5} x\cup y\cup z\cup w_2(B).
	\label{}
\end{equation}
Here $w_2(B)\in H^2[M_5, \Z_2]$ is the Stieffel-Whitney class of the SO(3) bundle $B$.

Let us see how this anomaly can be resolved by a $\U$ gauge theory. Notice that $\rho(\mb{g})=\mathds{1}$ for $\mb{g}\in \mathrm{SO}(3)$ because SO(3) is connected. For translations, let us for simplicity assume that $T_{x,y,z}$ act on the charges in the same way, denoted by $\rho_1$:
\begin{equation}
	\rho(T_x^{n_x}T_y^{n_y}T_z^{n_z})=\rho_1^{n_x+n_y+n_z}.
	\label{}
\end{equation}
This is natural if the cubic rotation symmetry is preserved. As shown in Sec. \ref{sec: anomaly formula}, there are three possibilities of how translation is associated with the duality transformation of a $\U$ gauge theory: the translation acts as the identity, the charge conjugation, or the $T$ transformation.

First we present an argument to rule out $\rho_1=\mathds{1}$.
We calculate the fractionalization classes using the K\"unneth decomposition:
\begin{equation}
	\H^2[\mathrm{SO}(3)\times \Z^3, \U]=\Z_2\times\U^3.
	\label{}
\end{equation}
The first factor of the above equation indicates that charges can transform as spin-$1/2$'s under SO(3). 
The $\U^3$ factor represents magnetic translation algebra in $xy, yz$ or $zx$ planes. However, we should notice that each of these $\U$ phase factors is a continuously tunable phase factor, and therefore should not form distinct fractionalization classes. This is similar to theta terms in topological response.\footnote{Mathematically, this is the distinction between deformation classes and not just isomorphism classes.} We conclude that when $\rho_1=\mathds{1}$ the fractionalization class of the translation symmetry is completely trivial, and thus can not happen in the presence of LSM anomaly, and, as a result, the translation must be mapped to a nontrivial element in the duality group. 

Next let us consider $\rho_1$ being the charge conjugation. In this case, we find
\begin{equation}
	\H^2_\rho[\Z^3, \U]=\Z_2.
	\label{}
\end{equation}
So there is only one nontrivial translation symmetry fractionalization pattern. An invariant that characterizes the fractionalization class is
\begin{equation}
	\frac{\eta(T_x,T_y)}{\eta(T_y, T_x)}\frac{\eta(T_y,T_z)}{\eta(T_z, T_y)}\frac{\eta(T_z,T_x)}{\eta(T_x, T_z)}=\pm 1.
	\label{}
\end{equation}

To resolve the LSM anomaly, clearly one of $\E$ and $\M$ has to carry spin-$1/2$, because the ``background matter fields" carry spin-1/2.  Without loss of generality, let $\E$ carry spin-$1/2$. It is natural to expect that $\M$ needs to carry the nontrivial translation symmetry fractionalization. We show in Appendix \ref{app: LSM anomaly} that this symmetry fractionalization pattern indeed realizes the LSM anomaly correctly. In contrast, the LSM anomaly cannot be realized if none of $\E$ and $\M$ carries spin-1/2, or none of them carries the nontrivial translation fractionalization pattern. The general condition for a $\U$ QSL to satisfy the LSM constraint due to these symmetries is given by  Eq. \eqref{eq: LSM matching equation}.

Let us list the possible symmetry-enriched $\U$ QSLs that can be realized in a lattice with spin-$1/2$ per unit cell. As before, we denote the one with spin-$1/2$ as $\E$, the spinon. Then $\M$ must carry integer spin, otherwise the state suffers from the SO(3) anomaly. There are only two types of $\U$ QSLs that satisfy the LSM constraint: $(\E_{b\frac{1}{2}}\M_{b\text{trn}})_-$ and $(\E_{b\frac{1}{2}\text{trn}}\M_{b\text{trn}})_-$, where `$()_-$' means that the translation symmetry acts as charge conjugation, $b\frac{1}{2}$ means a spin-1/2 boson, and $b\trn$ means a boson with nontrivial translation fractionalization. In Appendix \ref{app: parton}, we show that both of them can indeed be realized by explicit parton constructions.

On the other hand, if the lattice has an integer spin per unit cell, then the possible symmetric $\U$ QSLs are $(\E_b\M_b)_-, (\E_{b\frac{1}{2}}\M_b)_-, (\E_{b\frac{1}{2}\text{trn}}\M_b)_-, (\E_{b\text{trn}}\M_b)_-$ and $(\E_{b\text{trn}}M_{b\text{trn}})_-$.

Lastly, we consider the possibility that $\rho_1$ is realized as $\mb{T}^n$ for some nonzero integer $n$. Leaving a general classification of this case for future work, here we will briefly describe an example where one of the translations, say $T_z$, is mapped to $\mathbf{T}^2$. To this end, we use a fermionic parton construction to write the spin operator in terms of Abrikosov fermions: $\vec{S}_i=\frac{1}{2}f_i^\dag \vec\sigma f_i$, with the local gauge constraint $f_i^\dag f_i=1$ imposed. We then put the fermions into the a mean-field state described by a non-interacting Hamiltonian. The original spin system is recovered by coupling the fermions to $\U$ gauge field. For our purpose, we choose the following mean-field band structure: for all fermions on a given $xy$ plane, we make $f_\uparrow$ and $f_\downarrow$ both have the same Chern band with Chern number $C=1$. Together they form a $C=2$ band, which is the minimal required by the SU(2) spin symmetry (see Appendix \ref{app: Hall}). In this case, the translations do not change the fermionic gauge charge. Following a similar discussion in Sec. \ref{sec:H4}, it is easy to see that $T_z$ implements the $\mb{T}^2$ transformation. 

It is also possible to have a $\U$ QSL where all three translations act as $\mb{T}^2$. To construct such a state, one just needs to take three copies of the above state and make them rotationally symmetric, and turn on hybridization between the charges in these three $\U$ gauge theories. The resulting theory is an SO(3) and translation symmetric $\U$ gauge theory with an odd number of spin-1/2's per unit cell, in which translations in all three directions act as $\mb{T}^2$.

\subsection{Fermionic insulators}
\label{sec:fermionspt}

As the final application of our results, we study an example of interacting fermionic topological insulator protected by a unitary symmetry $G$~\cite{GuWen, ChengPRX2018}. For simplicity, we assume fermions transforming linearly under the symmetry group $G$, and $\rho(\mb{g})=\mathds{1}$ for $\mb{g}\in G$. After gauging the U(1) symmetry, one obtains a  U(1) gauge theory with fermionic gauge charges. A topologically nontrivial insulator can have magnetic monopoles carrying projective representation under $G$, provided that there is no 't Hooft anomaly in the gauged theory. 
 
To compute the anomaly, we first apply a $\mb{T}$ transformation so that the electric charge is bosonic. In other words, we may view the fermionic topological insulator as the result of ``ungauging" the $(1,1)$ dyon in a U(1) gauge theory with bosonic electric charge. Since $\rho$ is trivial, both $\nu$ and $\omega_\M$ are elements of $\H^2[G, \U]$. Because we assume that the fermion $(1,1)$ transforms linearly, it follows that $\nu=\omega_\M^{-1}$.

In the following we specify to an example with $G=\Z_{N_1}\times\Z_{N_2}$.
Projective representations of $G$ are classified according to $\H^2[\mathbb{Z}_{N_1}\times\mathbb{Z}_{N_2},\U]=\Z_{N_{12}}$, where $N_{12}$ is the greatest common divider of $N_1$ and $N_2$. We have the following explicit expressions for the $2$-cocycles:
\begin{equation}
	\omega(a,b) = \frac{2\pi p}{N_{12}} a_1 b_2,
	\quad
	p=0,1,\dots, N_{12}-1.
	\label{}
\end{equation}

Now let us analyze the obstruction class. Kunneth formula gives $\H^5[\Z_{N_1}\times\Z_{N_2},\U]=\Z_{N_1}\times\Z_{N_2}\times\Z_{N_{12}}^2$. It is clear that we just need to consider the $\Z_{N_{12}}^2$ part.  Ref. [\onlinecite{ChengPRX2018}] found a complete set of invariants, $e^{i\Omega_1},e^{i\Omega_2}, e^{i\Omega_{12}},e^{i\Omega_{21}}$, for cohomology classes. We review the definitions in Appendix \ref{app:math}. A straightforward calculation yields
\begin{equation}
\begin{split}
   &\Omega_{12}=-\pi p^2 \frac{N^{12}(N^{12}-1)N_2}{N_{12}^2},\\ &\Omega_{21}=-\pi p^2\frac{N^{12}(N^{12}-1)N_1}{N_{12}^2}.
\end{split}
\end{equation}
Here $N_{12}$ is the greatest common divisor of $N_1$ and $N_2$, and $N^{12}$ is the least common multiplier.
The obstruction class is trivial if and only if $e^{i\Omega_{12}}=e^{i\Omega_{21}}=1$.

For $N_1=N_2=N$, both of them reduce to $\pi{p^2(N-1)}$. The obstruction class is $e^{i\Omega_{12}}=(-1)^{p (N-1)}$, which is trivial for all odd $N$. For $N=2, p=1$ the obstruction class is nontrivial, which is the claim in Sec. \ref{sec:so3}. We conclude that there exists topologically nontrivial fermionic insulators protected by $\Z_N\times\Z_N$ symmetry for odd $N$.

Consider another family of examples, with $N_1=2^{n_1}, N_2=2^{n_2}$. Without loss of generality we assume $n_1\leq n_2$. The invariants are evaluated to 
\begin{equation}
\begin{gathered}
    \Omega_{12}=\pi p^2 2^{2(n_2-n_1)}(2^{n_2}-1),\\ \Omega_{21}=\pi p^2 2^{n_2-n_1}(2^{n_2}-1).
    \end{gathered}
\end{equation}
As long as $n_2>n_1$, the obstruction class always vanishes. The simplest example is $N_1=2, N_2=4$. In this fermionic SPT phase, a magnetic monopole carries a projective representation of $\Z_2\times\Z_4$. We notice that this state is  the same as the intrinsically interacting fermionic SPT phase found in Ref. \cite{ChengPRX2018}, which was obtained there essentially by using the group super-cohomology construction. It is worth mentioning that Ref. \cite{ChengPRX2018} only assumes the $\Z_2$ fermion parity conservation, which means that the U(1) charge conservation is not essential for the existence of this phase.

\section{Summary and discussion}
\label{sec: discussion}

In this work we have classified symmetry fractionalization and anomalies in a symmetry-enriched \spd{3} U(1) gauge theory with bosonic electric charges and a global symmetry group $G$, based on the conjecture that a $G$-symmetric $\U$ gauge theory can be viewed as a partially gauged SPT. We find that, in general, a symmetry-enrichment pattern is specified by 4 pieces of data: $\rho$, a map from $G$ to the SL(2, $\Z$) duality group which physically encodes how the symmetry permutes the fractional excitations, $\nu\in\H^2_{\rho\cdot s}[G, \U]$, the symmetry actions on the electric charge, $p\in\H^1_s[G, \Z]$, indication of certain domain wall decoration with bosonic integer quantum Hall states, and a torsor $n$ over $\H^3_{\rho}[G, \Z]$,  the symmetry actions on the magnetic monopole.

However, certain choices of $(\rho, \nu, p, n)$ are not physically realizable, \ie they are anomalous. We find that there are two levels of anomalies. The first level of anomalies obstruct the fractional excitations being deconfined, thus are referred to as the deconfinement anomaly. States with these anomalies can be realized on the boundary of a \spd{4} long-range entangled state. The deconfinement anomalies are classified by $\H^3_\rho[G, \U]$. If a state does not suffer from a deconfinement anomaly, there can be still the second level of anomaly, the more familiar 't Hooft anomaly, which forbids certain types of symmetry fractionalization patterns.  States with these anomalies can be realized on the boundary of a \spd{4} short-range entangled state. These 't Hooft anomalies are classified by $\H^5_s[G, U(1)]$.

We have applied these results to some interesting physical examples. Besides being able to reproduce and extend the previous results in Ref. \cite{Zou2017}, we also utilized our anomaly formula to study the LSM-type constraints on a $\U$ QSL, and some interesting interacting fermionic topological insulators.

Below we briefly discuss some future directions.

One class of U(1) QSLs left out from our classification are those with $\theta=\pi$ in the presence of anti-unitary symmetries, and more generally U(1) gauge theories with fermionic electric charge. To extend our approach to these cases, it is necessary to have a complete understanding of interacting fermionic insulators.

We have briefly mentioned the possibility that certain unitary infinite-order symmetries, such as translations, can be realized as modular transformations, corresponding to a nonzero $[p]\in \H^1[G, \Z]$. We have demonstrated the possible $\H^3$ deconfinement obstruction class in these states. A more complete study of such phases, as well as their potential relation with the fractonic phases, will be left for future work.

Our classification principle only allows global unitary symmetries to act as the identity, charge conjugation or modular transformations in the duality group. An interesting open question is: to what extent a global symmetry acting as the $\mb{S}$-duality transformation, for example, is anomalous, and what is the nature of the anomaly if there is any? We note that there have been a few works on $\U$ gauge theories with a global symmetry realized as the $\mb{S}$-duality ~\cite{KravecPRL2013, Bi_arxiv2015, Cordova_arxiv}. In some cases the $\U$ gauge theory is actually the ``all-fermion" one, which is the boundary of a \spd{4} invertible topological phase~\cite{Kapustin2014, wang2014}. We will leave this for future investigations.

Many of our results can be generalized to a $\Z_N$ gauge theory in a straightforward manner. In particular, the parametrization of 4-cocycles can be applied to the $\Z_N$ case without much modifications. Physically, however, the magnetic excitations are now extended loop-like objects. It will be important to develop a physical understanding of symmetry fractionalization on loop-like excitations, which will be addressed in future publications.

\section{Acknowledgements}

L. Z. and M.C. would like to thank T. Senthil and Chong Wang for discussions. M. C. would like to thank Chao-Ming Jian for many enlightening conversations, and Yang Qi for help on group cohomology calculations.
S.Q.N. is  supported by NSFC (Grant Nos.11574392, 11574172), the Ministry of Science and Technology of China (Grant No. 2016YFA0300504).
L. Z. is supported by NSF grant DMR-1608505 and the John Bardeen Postdoctoral Fellowship at Perimeter Institute. Research at Perimeter Institute is supported in part by the Government of Canada through the Department of Innovation, Science and Economic Development Canada and by the Province of Ontario through the Ministry of Colleges and Universities. DC is supported by faculty startup funds at Cornell University. M. C. is supported by Alfred P. Sloan Research Fellowship and NSF CAREER (DMR-1846109).

\textbf{Note added:} while the manuscript was being finalized, a preprint on closely related topic appeared on arXiv~\cite{otherpaper}. We are also aware of a related work by Xu Yang and Ying Ran~\cite{XuRan}.

\bibliography{references.bib}

\clearpage
\onecolumngrid
\appendix

\section{Some useful mathematical results on group cohomology} \label{app:math}
In this appendix we collect a few mathematical results used in this work.

\subsection{Review of group cohomology}
In this section, we provide a brief review of group cohomology for finite groups.
Given a finite group $G$, let $M$ be an Abelian group equipped with a $G$ action $\rho: G \times M \rightarrow M$, which is compatible with group multiplication. In particular, for any $\mathbf{g}\in G$ and $a,b \in M$, we have
\begin{equation}
\rho_\mathbf{g}(ab)=\rho_\mathbf{g}(a) \rho_\mathbf{g}(b).
\label{}
\end{equation}
(We leave the group multiplication symbols implicit.) Such an Abelian group $M$ with a $G$ action $\rho$ is called a $G$-module.

Let $\omega(\mathbf{g}_1, \dots,\mathbf{g}_n)\in M$ be a function of $n$ group elements $\mathbf{g}_j \in G$ for $j=1,\dots,n$. Such a function is called an $n$-cochain, and the set of all $n$-cochains is denoted as $C^n[G, M]$. They naturally form a group under multiplication,
\begin{equation}
  (\omega\cdot\omega')(\mb{g}_1, \dots, \mb{g}_n)=\omega(\mb{g}_1, \dots, \mb{g}_n)\omega'(\mb{g}_1, \dots, \mb{g}_n),
  \label{}
\end{equation}
and the identity element is the trivial cochain $\omega(\mb{g}_1,\dots,\mb{g}_n)=1$.

We now define the ``coboundary'' map $\delta: C^n[G, M] \rightarrow C^{n+1}[G, M]$ acting on cochains to be
\begin{equation}
\begin{split}
\delta\omega  (\mathbf{g}_1,\dots,\mathbf{g}_{n+1})= \rho_{\mathbf{g}_1}&[\omega(\mathbf{g}_2,\dots,\mathbf{g}_{n+1})] 
	\times \prod_{j=1}^n \omega^{(-1)^j}(\mathbf{g}_1,\dots,\mathbf{g}_{j-1},\mathbf{g}_j\mathbf{g}_{j+1},\mathbf{g}_{j+2},\dots,\mathbf{g}_{n+1}) \\
    &\times \omega^{(-1)^{n+1}}(\mathbf{g}_1,\dots,\mathbf{g}_{n})
.
\end{split}
\label{}
\end{equation}
One can directly verify that $\delta^2\omega=1$ for any $\omega \in C^n[G, M]$, where $1$ is the trivial cochain in $C^{n+2}[G, M]$. 

With the coboundary map, we next define $\omega\in C^n[G, M]$ to be an $n$-cocycle if it satisfies the condition $\delta\omega=1$. We denote the set of all $n$-cocycles by
\begin{equation}
\begin{split}
Z^n_{\rho}[G, M]
 = \{ \, \omega\in C^n[G, M] \,\, | \,\, \delta\omega=1 \, \}.
\end{split}
\label{}
\end{equation}
We also define $\omega\in C^n[G, M]$ to be an $n$-coboundary if it satisfies the condition $\omega= \delta \mu $ for some $(n-1)$-cochain $\mu \in C^{n-1}[G, M]$. We denote the set of all $n$-coboundaries by $ B^n_{\rho}(G, M)$.
Namely,
\begin{equation}
\begin{split}
 B^n_{\rho}(G, M) 
=\{ \, \omega\in C^n[G, M] \,\, | \,\, \exists \mu \in C^{n-1}[G, M] : \omega = \delta\mu \, \}
.
\end{split}
\label{}
\end{equation}

Clearly, $B^n_{\rho}[G, M] \subset Z^n_{\rho}[G, M] \subset C^n[G, M]$. In fact, $C^n$, $Z^n$, and $B^n$ are all groups and the co-boundary maps are homomorphisms. It is easy to see that $B^n_{\rho}[G, M]$ is a normal subgroup of $Z^n_{\rho}[G, M]$. Since $\delta$ is a boundary map, we think of the $n$-coboundaries as being trivial $n$-cocycles, and it is natural to consider the quotient group
\begin{equation}
	\H^n_{\rho}[G, M]=\frac{Z^n_{\rho}[G, M]}{B^n_{\rho}[G, M]}
,
\label{}
\end{equation}
which is called the $n$-th group cohomology. In other words, $\mathcal{H}^n_{\rho}[G, M]$ collects the equivalence classes of $n$-cocycles that only differ by $n$-coboundaries.

The algebraic definition we give for group cohomology is most convenient for discrete groups. For continuous group, formally the same definition applies but one has to impose proper continuity conditions on the cocycle functions.

\subsection{$\H^n[G, \U]$ and $\H^{n+1}[G, \Z]$}
\label{appendix:mapping}

The equivalence of the two cohomology groups follow from the short exact sequence $0\rightarrow \Z\rightarrow\mathbb{R}\rightarrow \U\rightarrow 0$, and the fact that $\H^n[G, \mathbb{R}]=\Z_1$ for compact/finite groups. Below we write down the explicit mapping between cocycles.

Given a $n$-cocycle $[\omega]\in \H^n[G, \U]$, we define $\omega=e^{i\hat{\omega}}$ where $\hat{\omega}\in \mathbb{R}$. $\delta \omega=1$ translates to $\delta \hat{\omega}\in 2\pi\Z$, where $\delta$  for $\hat{\omega}$ is defined additively. We can now define a $(n+1)$-cocycle $\nu\in Z^{n+1}[G, \Z]$ as
	\begin{equation}
		\nu=\frac{1}{2\pi}\delta \hat{\omega}.
		\label{eqn:bock}
	\end{equation}
	If we shift $\nu$ by a coboundary $\delta \mu$ where $\mu\in C^n[G, \Z]$, it simply amounts to shifting $\hat{\omega}\rightarrow \hat{\omega}+2\pi\mu$, which does not affect the value of $\omega$.

	On the other hand, if we change $\omega$ by a coboundary, or equivalently $\hat{\omega}\rightarrow \hat{\omega}+\delta\hat{\varepsilon}$ where $\hat{\varepsilon}\in C^{n-1}[G, \mathbb{R}/2\pi\Z]$, we find that the corresponding $\nu$ remains the same since $\delta^2\hat{\varepsilon}=0$.

	Eq. \eqref{eqn:bock} is a special case of the Bockstein homomorphism, and commonly written as $\frac{1}{2\pi}\beta\omega$ for $[\omega]\in\H^{n}[G, \mathbb{R}/2\pi\Z]$ (without explicitly specifying the lift from $\mathbb{R}/2\pi\Z$ to $\mathbb{R}$).

\subsection{Slant products}
\label{sec:slant}

A $k$-slant product maps a $n$-cochain $\omega_n$ to a $(n-k)$-cochain $\omega_{n-k}$. If $\omega_n$ is a $n$-cocycle, generally $\omega_{n-k}$ is not a cocycle except for $k=1$. However, if $i_g\omega_{n}$ is a $(n-1)$-coboundary, then $i_{g,h}\omega_n$ is a $(n-2)$-cocycle. For more details, see Ref. \cite{NatPRB2017}. Notice that 1-slant product is often just called slant product.

Now we give the general definition of $1$-slant product, often just known as the slant product. Let us consider $M$ being a $G$-module with trivial action, and $\mb{g}\in G$ be an arbitrary element. We define the 1-slant product $i_\mb{g}: C^n[G, M]\rightarrow C^{n-1}[G, M]$:
\begin{equation}
 i_\mathbf{g}\omega (\mb{g}_1,\ldots,\mb{g}_{n-1})
  = \prod_{j=0}^{n-1}\omega(\mb{g}_1,\ldots, \mb{g}_{j},\mb{g},\mb{g}_{j+1},\ldots,\mb{g}_{n-1})^{(-1)^{n-1+j}}.
\quad \label{eq:slant_product}
\end{equation}
It can be shown that $\delta(i_\mb{g}\omega)=i_\mb{g}(\delta\omega)$. Therefore, $i_\mb{g}$ is in fact a group homomorphism:
\begin{equation}
	i_\mb{g}: \mathcal{H}^n[G, M]\rightarrow \mathcal{H}^{n-1}[G, M].
	\label{}
\end{equation}

For example, for a 5-cochain $\omega_5$, the 1-slant product is defined as
\begin{align}
i_a\omega_5(\mb{g,h,k,l})=\frac{\omega_5(a, \mb{g,h,k,l})\omega_5(\mb{g,h},a,\mb{k,l})\omega_5(\mb{g,h,k,l},a)}{\omega_5( \mb{g},a,\mb{h,k,l})\omega_5(\mb{g,h,k},a,\mb{l})},
\end{align}
The 2-slant product is given by
 \begin{align}
(i_\mb{g,h} {\omega_5})(a,b,c)=\frac{\omega_5(a,b,\mb{g},c,\mb{h})\omega_5(a,\mb{g},b,\mb{h},c)\omega_5(\mb{g},a,\mb{h},b,c)\omega_5(\mb{g},a,b,c,\mb{h})}{\omega_5(a,b,c, \mb{g,h})\omega_5(a,b,\mb{g,h},c)\omega_5(a,\mb{g,h},b,c)\omega_5(\mb{g,h},a,b,c)\omega_5(a,\mb{g},b,c,\mb{h})\omega_5(\mb{g},a,b,\mb{h},c) }
\end{align}
and 3-slant product is given by
\begin{align}
	(i_{a,b,c} {\omega_5})(\mb{g,h})=[(i_\mb{g,h}\omega_5)(a,b,c)]^{-1}
\end{align}

\subsection{Invariants for $\H^5[\Z_{N_1}\times\Z_{N_2}, \U]$}
Let $\mb{e}_i$ be the generator of the $\Z_{N_i}$ subgroup. Define the following invariants
\begin{equation}
    e^{i\Omega_i}=\prod_{m,n=1}^{N_i}i_{\mb{e}_i}\nu(\mb{e}_i, m\mb{e}_i, \mb{e}_i, n\mb{e}_i).
\end{equation}
And, for $i\neq j$,
\begin{equation}
    e^{i\Omega_{ij}}=\prod_{m=1}^{N^{ij}}\prod_{n=1}^{N_i}i_{\mb{e}_j, m\mb{e}_j}\nu(\mb{e}_i, n\mb{e}_i, \mb{e}_i)i_{\mb{e}_j}\nu(\mb{e}_i, n\mb{e}_i, \mb{e}_i, m\mb{e}_i).
\end{equation}
Ref. [\onlinecite{ChengPRX2018}] showed that these are a complete set of invariants for cohomology classes in $\H^5[\Z_{N_1}\times\Z_{N_2}, \U]$.

\onecolumngrid

\section{Parametrization of 4-cocycles and classification of anomalies}
\label{app:4cocycle}

In this appendix, we present the detailed derivation of the structure of the cohomology group $\H^4[\mathcal{G}, \U]$ given in Sec. \ref{sec:H4}.

To be self-contained, we first repeat the reasoning leading to the results here. The bulk properties of a symmetry-enriched U(1) gauge theory is specified by the symmetry fractionalization patterns of the global symmetry $G$ on the electric and magnetic charges. However, not all symmetry fractionalization patterns can be physically realized, and our goal is to obtain a set of sufficient and necessary conditions under which a symmetry fractionalization pattern is anomaly-free.

To do so, we will utilize that the symmetry-enriched U(1) gauge theory with global symmetry $G$ can be viewed as a gauged bosonic SPT phase protected by a symmetry $\mc{G}$, a $\U$ extension of $G$. Such an group extension $\mc{G}$ is specified by a 2-cocycle in $[\nu]\in\H^2_{\rho\cdot s}[G, \U]$, and it physically encodes the symmetry fractionalization pattern of the electric charge of this symmetry-enriched $\U$ gauge theory. As shown in Sec. \ref{sec: gauged SPT}, the relevant bosonic SPT phases can all be obtained from group cohomology, and each of them is specified by a 4-cocycle in $\H^4_{\rho,s}[\mc{G}, U(1)]$. Below, from this 4-cocycle, we will extract the data of the projective representation of the dual magnetic charge under $G$. We will also formulate a set of sufficient and necessary conditions for the symmetry fractionalization patterns to be anomaly-free.

A 4-cocycle in $\H^4_{\rho,s}[\mc{G}, \U]$ can be represented by a $\U$-valued function of 4 elements in group $\mc{G}$. Using $a, b, c, d,\cdots\in \mathbb{R}/2\pi\Z$ to denote  elements in $\U$, and $\mb{g}, \mb{h}, \mb{k}, \mb{l},\cdots\in G$ to denote elements in $G$. An element in $\mc{G}$ can be denoted by $a_{\mb{g}}$, which means this element can be viewed as a composite of $a$ and $\mb{g}$. A 4-cocycle can be written as $\omega(a_{\mb{g}}, b_{\mb{h}}, c_{\mb{k}}, d_{\mb{l}})\in\U$, which satisfies the following 4-cocycle equation:
\beq \label{eq: 4-cocycle equation}
\frac{\omega^{s(\mb{g})}(b_{\mb{h}}, c_{\mb{k}}, d_{\mb{l}}, e_{\mb{m}})\omega(a_{\mb{g}}, b_{\mb{h}}\times c_{\mb{k}}, d_{\mb{l}}, e_{\mb{m}})\omega(a_{\mb{g}}, b_{\mb{h}}, c_{\mb{k}}, d_{\mb{l}}\times e_{\mb{m}})}{\omega(a_{\mb{g}}, b_{\mb{h}}, c_{\mb{k}}, d_{\mb{l}})\omega(a_{\mb{g}}\times b_{\mb{h}}, c_{\mb{k}}, d_{\mb{l}}, e_{\mb{m}})\omega(a_{\mb{g}}, b_{\mb{h}}, c_{\mb{k}}\times d_{\mb{l}}, e_{\mb{m}})}=1
\eeq
where the group multiplication law is
\beq
a_{\mb{g}}\times b_{\mb{h}}=[a+ ^\mb{g}b+\nu(\mb{g}, \mb{h})]_{\mb{gh}}
\eeq
with $^\mb{g}b=\rho(\mb{g})\cdot s(\mb{g})\cdot b$, and the two $\Z_2$ gradings are defined as
\beq
s(\mb{g})=
\left\{
\begin{array}{lr}
1, & \mb{g}{\rm\ is\ unitary}\\
-1, & {\rm otherwise}
\end{array},
\right.
\rho(\mb{g})=
\left\{
\begin{array}{lr}
-1, & \mb{g}{\rm\ contains\ charge\  conjugation,}\\
1, & {\rm otherwise.}
\end{array}
\right.
\eeq
and $\nu(\mb{g}, \mb{h})\in\H^2_{\rho,s}[G, \mathbb{R}/2\pi\Z]$.

A 4-cocycle has a gauge freedom, which states that $\omega(a_{\mb{g}}, b_{\mb{h}}, c_{\mb{k}}, d_{\mb{l}})$ is physically equivalent to
\beq
\omega(a_{\mb{g}}, b_{\mb{h}}, c_{\mb{k}}, d_{\mb{l}})\cdot\frac{u^{s(\mb{g})}(b_{\mb{h}}, c_{\mb{k}}, d_{\mb{l}})u(a_{\mb{g}}, b_{\mb{h}}\times c_{\mb{k}}, d_{\mb{l}})u(a_{\mb{g}}, b_{\mb{h}}, c_{\mb{k}})}{u(a_{\mb{g}}\times b_{\mb{h}}, c_{\mb{k}}, d_{\mb{l}})u(a_{\mb{g}}, b_{\mb{h}}, c_{\mb{k}}\times d_{\mb{l}})}
\eeq
where $u(a_{\mb{g}}, b_{\mb{h}}, c_{\mb{k}})$ is $\U$-valued.

Below we will first derive a general parametrization of the 4-cocycles $\omega(a_{\mb{g}}, b_{\mb{h}}, c_{\mb{k}}, d_{\mb{l}})$, and find the anomaly-free conditions.

\subsection{Parameterize the 4-cocycles}

For generality of the calculations, throughout this section we use $A=\U$ to denote the gauge group.

The 4-cocycles contain much redundant information due to gauge freedom. To have a useful form of the 4-cocycles, we can use the gauge freedom to fix some of them. In particular, we will fix all of the following 4-cocycles to be 1:
\beq \label{eq: gauge fixing 4-cocycle}
\begin{split}
1&=\omega(a, b, c, d)\\
&=\omega(0, a_{\mb{g}}, b_{\mb{h}}, c_{\mb{k}})=\omega(a_{\mb{g}}, 0, b_{\mb{h}}, c_{\mb{k}})=\omega(a_{\mb{g}}, b_{\mb{h}}, 0, c_{\mb{k}})=\omega(a_{\mb{g}}, b_{\mb{h}}, c_{\mb{k}}, 0)\\
&=\omega(0_{\mb{g}}, b, c, d)=\omega(a, 0_{\mb{g}}, b, c)=\omega(a, b, 0_{\mb{g}}, c)\\
&=\omega(a, 0_{\mb{g}}, 0_{\mb{h}}, b)=\omega(0_{\mb{g}}, a, 0_{\mb{h}}, b)=\omega(0_{\mb{g}}, a, b, c_{\mb{h}})=\omega(a_{\mb{g}}, 0_{\mb{h}}, b, c)=\omega(a, b, 0_{\mb{g}}, 0_{\mb{h}})\\
&=\omega(0_{\mb{g}}, 0_{\mb{h}}, 0_{\mb{k}}, a)=\omega(0_{\mb{g}}, 0_{\mb{h}}, b, c_{\mb{k}})=\omega(0_{\mb{g}}, a, 0_{\mb{h}}, b_{\mb{k}})=\omega(a_{\mb{g}}, 0_{\mb{h}}, b, c_{\mb{k}})
\end{split}
\eeq
Notice that $\omega(a,b,c,d)=1$ can always be done because $\H^4[A,\U]=\Z_1$, which is true for $A=\Z_N$ as well. For a general Abelian gauge theory, this condition means that it is untwisted. 

We will then express a general 4-cocycle $\omega(a_{\mb{g}}, b_{\mb{h}}, c_{\mb{k}}, d_{\mb{l}})$ in terms of the following objects:
\begin{equation}
\begin{split}
    \omega_{\mb{g,h,k}}(a)&\equiv \omega(a,0_\mb{g},0_\mb{h},0_\mb{k}),\\
    \omega_{\mb{g,h}}(a,b)&\equiv \omega(a,b,0_\mb{g},0_\mb{h})\\
    \omega_{\mb{g}}(a,b,c)&\equiv \omega(a,b,c,0_\mb{g})
    \end{split}
    \label{eqn:data}
\end{equation}

After a rather tedious calculation, by applying the 4-cocycle equations Eq. (\ref{eq: 4-cocycle equation}) for various group elements, one can show that after the above gauge fixing the 4-cocycle can be written as
\beq \label{eq: parameterize 4-cocycles}
\begin{split}
\omega(a_{\mb{g}}, b_{\mb{h}}, c_{\mb{k}}, d_{\mb{l}})
=&
\omega(0_{\mb{g}}, 0_{\mb{h}}, 0_{\mb{k}}, 0_{\mb{l}})\omega^{s(\mb{g})}_{\mb{h}, \mb{k}, \mb{l}}(^{\overline{\mb{g}}}a) 
		\omega_{\mb{k,l}}^{s(\mb{gh})}(\ga{gh}{a},\ga{h}{b})\omega_{\mb{k,l}}^{s(\mb{gh})}\big(\ga{gh}{\nu(\mb{g,h})},\g{h}(\ga{g}{a}+b)\big)
		\cdot\alpha(a_{\mb{g}}, b_{\mb{h}}, c_{\mb{k}}, d_{\mb{l}})
\end{split}
\eeq
with
\beq
\begin{split}
\alpha(a_{\mb{g}}, b_{\mb{h}}, c_{\mb{k}}, d_{\mb{l}})=&\frac{\omega_{\mb{l}}^{s(\mb{ghk})}(^{\overline{\mb{ghk}}}(a+\nu(\mb{g}, \mb{hk})), ^{\overline{\mb{hk}}}(b+\nu(\mb{h}, \mb{k})), ^{\overline{\mb{k}}}c)}{\omega^{s(\mb{gh})}_{\mb{k}}(^{\overline{\mb{gh}}}\nu(\mb{g}, \mb{h}), ^{\overline{\mb{gh}}}a, ^{\overline{\mb{h}}}b)}
\cdot
\frac{\omega_{\mb{l}}^{s(\mb{ghk})}(^{\overline{\mb{ghk}}}a, ^{\overline{\mb{hk}}}\nu(\mb{h}, \mb{k}), ^{\overline{\mb{hk}}}b)}{\omega_{\mb{l}}^{s(\mb{ghk})}(^{\overline{\mb{hk}}}\nu(\mb{h}, \mb{k}), ^{\overline{\mb{ghk}}}a, ^{\overline{\mb{hk}}}b)}\\
\cdot&\frac{\omega_{\mb{l}}^{s(\mb{ghk})}(^{\overline{\mb{ghk}}}\nu(\mb{g}, \mb{hk}), ^{\overline{\mb{hk}}}\nu(\mb{h}, \mb{k}), ^{\overline{\mb{hk}}}(^{\overline{\mb{g}}}a+b))}{\omega_{\mb{l}}^{s(\mb{ghk})} (^{\overline{\mb{ghk}}}\nu(\mb{g}, \mb{hk}), ^{\overline{\mb{ghk}}}a, ^{\overline{\mb{hk}}}(b+\nu(\mb{h}, \mb{k})))}
\cdot\frac{\omega_{\mb{kl}}^{s(\mb{gh})}(^{\overline{\mb{gh}}}\nu(\mb{g}, \mb{h}), ^{\overline{\mb{gh}}}a, ^{\overline{\mb{h}}}b)}{\omega_{\mb{l}}^{s(\mb{ghk})}(^{\overline{\mb{ghk}}}\nu(\mb{gh}, \mb{k}), ^{\overline{\mb{ghk}}}\nu(\mb{g}, \mb{h}), ^{\overline{\mb{hk}}}(^{\overline{\mb{g}}}a+b))}.
\end{split}
\eeq

The data defined in Eq. \eqref{eqn:data} satisfies a number of consistency relations following from the cocycle condition. We now explain these relations, which also help uncover their physical interpretations:
\begin{enumerate}
	\item For a fixed $\mb{g}$, $\omega_\mb{g}(a,b,c)$ is a $3$-cocycle on $A$. Namely,
	\beq
	\frac{\omega_{\mb{g}}(b, c, d)\omega_{\mb{g}}(a, bc, d)\omega_{\mb{g}}(a, b, c)}{\omega_{\mb{g}}(ab, c, d)\omega_{\mb{g}}(a, b, cd)}=1
	\eeq
	\item The cohomology classes $[\omega_\mb{g}]$ satisfy $[\omega_\mb{g}]\cdot [\omega_\mb{h}^{s(\mb{g})}]=[\omega_\mb{gh}]$. More precisely:
		\begin{equation}
			\frac{\omega_\mb{g}(a,b,c)\omega^{s(\mb{g})}_\mb{h}(\g{g}a,\g{g}b,\g{g}c)}{\omega_\mb{gh}(a,b,c)}=(\delta \omega_\mb{g,h})(a,b,c).
			\label{eqn:4-cocycle-1}
		\end{equation}
		Here
		\begin{equation}
		    (\delta \omega)(a,b,c)=\frac{\omega(a,bc)\omega(b,c)}{\omega(a,b)\omega(ab,c)}.
		\end{equation}
	\item 
		\begin{equation}
			\frac{\omega_\mb{g,h}(a,b)\omega_\mb{gh,k}(a,b)}{\omega^{s(\mb{g})}_\mb{h,k}(\g{g}a,\g{g}b)\omega_\mb{g,hk}(a,b)}
			\frac{\omega_\mb{g,h,k}(ab)}{\omega_\mb{g,h,k}(a)\omega_\mb{g,h,k}(b)}=\bigg(
			\frac{\omega_\mb{k}(\g{gh}a, \g{gh}\nu(\mb{g,h}),\g{gh}b)}{\omega_\mb{k}(\g{gh}\nu(\mb{g,h}),\g{{g}h}a,\g{gh}b)\omega_\mb{k}(\g{gh}a, \g{gh}b,\g{gh}\nu(\mb{g,h}))}\bigg)^{s(\mb{gh})}.
			\label{eqn:4cyc-cond-2}
		\end{equation}
	\item
		\begin{equation}
			\begin{split}
			&\frac{\omega_\mb{g,h,k}(a)\omega_\mb{g,hk,l}(a) \omega^{s(\mb{g})}_\mb{h,k,l}(\g{g}a)}
			{\omega_\mb{gh,k,l}(a)\omega_\mb{g,h,kl}(a)} = \left(\frac{\omega_{\mb{k,l}}(\ga{gh}{a}, \ga{gh}{\nu(\mb{g,h}}))}{\omega_{\mb{k,l}}( \ga{gh}{\nu(\mb{g,h})}, \ga{gh}{a})
			}\right)^{s(\mb{gh})}
			\\
			&{\bigg(}\frac{\omega_\mb{l}(\g{gh{k}}a,\g{gh{k}}\nu(\mb{g,hk}),\g{h{k}}\nu(\mb{h,k}))\omega_\mb{l}(\g{ghk}\nu(\mb{g,hk}),\g{hk}\nu(\mb{h,k}),\g{ghk}a)\omega_\mb{l}(\g{gh{k}}\nu(\mb{gh,k}), \g{{g}h{k}}a, \g{gh{k}}\nu(\mb{g,h}))}
			{\omega_\mb{l}(\g{ghk}\nu(\mb{g,hk}),\g{ghk}a,\g{hk}\nu(\mb{h,k}))\omega_\mb{l}(\g{ghk}\nu(\mb{gh,k}),\g{{gh}k}\nu(\mb{g,h}),\g{ghk}a)\omega_\mb{l}(\g{gh{k}}a,\g{gh{k}}\nu(\mb{gh,k}),\g{gh{k}}\nu(\mb{g,h})}\bigg)^{-s(\mb{ghk})}
			\end{split}
			\label{heiheiha}
		\end{equation}
\end{enumerate}

The first two conditions imply that $\omega_\mb{g}$ defines a cohomology class in $\H^1_s[G, \H^3[A, \U]]=\H^1_s[G, \Z]$. We will fix
\begin{equation}
    \omega_\mb{g}(a,b,c)=\omega(a,b,c)^{p(\mb{g})}, 
\end{equation}
where $\omega(a,b,c)$ is a generating $3$-cocycle of $\H^3[A,\U]$. 
$p(\mb{g})\in\mathbb{Z}$ satisfies $p(\mb{g})+s(\mb{g})p(\mb{h})=p(\mb{gh})$. 
Without loss of generality, we choose
\beq \label{eq: well-known solution}
\omega(a, b, c)=\exp\left({i a\cdot\frac{[b]_{2\pi}+[c]_{2\pi}-[b+c]_{2\pi}}{2\pi}}\right),
\label{eqn:onedefect_3cocycle}
\eeq
where $[a]_{2\pi}\in[{0}, 2\pi)$ and $a=[a]_{2\pi}\ ({\rm mod\ } 2\pi)$.

Notice if $G$ is unitary and compact/finite, we can always set $p(\mb{g})=0$ because $\H^1[G, \Z]=\Z_1$. On the other hand, if $G$ contains time reversal symmetry, {then $\H^1_s[G,\Z]=\Z_2$. 

For technical reasons we introduce a 2-cochain $y_\mb{g}(a,b)$ of $A$ such that:
\begin{equation}
    \omega(\ga{g}{a}, \ga{g}{b},\ga{g}{c})=\omega(a,b,c) (\delta y_\mb{g})(a,b,c).
\end{equation}
An explicit choice for $y_\mb{g}(a,b)$ is given by
\begin{equation}
    y_\mb{g}(a,b)= \exp\left[ \frac{i}{2\pi} \big( \ga{g}{a}[\ga{g}{b}]_{2\pi}-a[b]_{2\pi}\big)\right].
\end{equation}
Also, note that, because $\H^2[A, \U]=\Z_1$, we can write the slant product of $\omega$ as a coboundary:
\begin{equation}
    (i_c\omega)(a,b)=\frac{\mu_c(a)\mu_c(b) }{\mu_c(a+b)}.
\end{equation}
Explicitly, we have
\begin{equation}
    \mu_a(b)= e^{\frac{i}{2\pi}[a]_{2\pi}[b]_{2\pi}}.
\end{equation}

Solving Eq. \eqref{eqn:4-cocycle-1} yields 
\begin{equation}
     \omega_\mb{g,h}(a,b)=y_\mb{g}^{-s(\mb{g})p(\mb{h})}(a,b).
\end{equation}

We proceed to solve Eq. \eqref{eqn:4cyc-cond-2}. Define
\begin{equation}
    \tilde{\omega}_\mb{g,h,k}(a)=
    \omega_{\mb{g,h,k}}(a)\mu(a,{\nu(\mb{g,h})})^{-s(\mb{gh})p(\mb{k})} \left(\frac{y_\mb{gh}(a,\nu(\mb{g,h}))}{y_\mb{gh}(\nu(\mb{g,h}),a)}\right)^{s(\mb{gh})p(\mb{k})}.
    \label{eqn:newomega}
\end{equation}
Eq. \eqref{eqn:4cyc-cond-2} can be written as
\begin{equation}
    \tilde{\omega}_{\mb{g,h,k}}(a)\tilde{\omega}_{\mb{g,h,k}}(b)=\tilde{\omega}_{\mb{g,h,k}}(ab),
\end{equation}
which means we can write
\begin{equation}
    \tilde{\omega}_{\mb{g,h,k}}(a)=e^{ian(\mb{g,h,k})},
\end{equation}
where $n(\mb{g,h,k})\in \Z$. Then Eq. \eqref{heiheiha} becomes
\begin{equation}
    e^{ia(\delta_\rho n)(\mb{g,h,k,l})}
    =\left[
    \frac{\mu_a(\nu(\mb{gh,k}))\mu_a(\nu(\mb{g,h}))}{\mu_a(\nu(\mb{g,hk}))\mu_a( {}^{\mb{g}}\nu(\mb{h,k}))}
    \frac{\mu_{\nu(\mb{gh,k})}(a)\mu_{\nu(\mb{g,h})}(a)}{\mu_{\nu(\mb{g,hk})}(a)\mu_{{}^{\mb{g}} \nu(\mb{h,k})}(a)}
    \cdot
    \frac{\mu(a,{}^{\mb{g}} \nu(\mb{h,k}))}{\mu(\ga{g}{a},{ \nu(\mb{h,k})})}
    \frac{y_\mb{g}({}^{\mb{g}}\nu(\mb{h,k}),a)}{y_\mb{g}(a,{}^{\mb{g}}\nu(\mb{h,k}))}
    \right]^{s(\mb{ghk})p(\mb{l})}.
\end{equation}

Using the explicit expressions for $y$ and $\mu$ given above, we obtain the following condition:
\begin{equation}
    (\delta_\rho n)(\mb{g,h,k,l})=\frac{1}{2\pi}
    2s(\mb{ghk})p(\mb{l})\Big(
     [  \nu(\mb{g,h})] + [  \nu(\mb{gh,h})]-[  \nu(\mb{g,hk})]- {}^{\mb{g}}[  \nu(\mb{h,k})]
    \Big).
\end{equation}
Note that this equality must hold \emph{exactly} as both sides are integers. To obtain this result we have used the fact that $\mb{g}$ acting on $A$ is either the identity or the conjugation, which applies to all cases studied in this paper.
Define $\gamma(\mb{g,h,k})=2s(\mb{gh})[\nu(\mb{g,h})]p(\mb{k})$, the above equation takes the form 
\begin{equation}
\delta_\rho n=\frac{1}{2\pi}\delta_\rho \gamma.
\label{eqn:eqforn}
\end{equation}
This means $\gamma$ must be a trivial $3$-cocycle in $\H^3[G, \mathbb{R}/2\pi \Z]$, otherwise there is no way to construct a $4$-cocycle out of the corresponding $[\nu]$ and $[p]$. We will refer to $[\gamma]$ as a $\H^3$ obstruction class.

Suppose the $\H^3$ obstruction class vanishes, then one can find solutions for $n$ from Eq. \eqref{eqn:eqforn}. Two solutions $n$ and $n'$ must satisfy $\delta_\rho(n-n')=0$, \ie they differ by an integer-valued $3$-cocycle of $G$. Therefore, in this case $n$ is classified by a torsor over $\H^3[G, \Z]$. As argued in the main text, this $n$ encodes the symmetry actions on the magnetic monopole, and it is related to $\omega_\M$ by $n=\frac{1}{2\pi}\delta\hat\omega_\M+n_0$, with $\omega_\M=e^{i\hat\omega_\M}$ and $n_0$ an integral 3-cochain satisfying $\delta_\rho n_0=\Gamma$, which is used as a ``reference" solution. In other words, starting from a particular solution $n_0$, we can construct a new one $n_0+\frac{1}{2\pi}\delta \hat{\omega}_\M$, in which the projective representation of the monopole is modified by $\omega_\M$ compared to the reference state.

Notice as long as $G$ is unitary, or $G$ contains time reversal with $\theta=0$, $\alpha(a_{\mb{g}}, b_{\mb{h}}, c_{\mb{k}}, d_{\mb{l}})=1$ and the 4-cocycle has a simple form
\beq \label{eq: parameterize 4-cocycles simpler}
\omega(a_{\mb{g}}, b_{\mb{h}}, c_{\mb{k}}, d_{\mb{l}})
=
\omega(0_{\mb{g}}, 0_{\mb{h}}, 0_{\mb{k}}, 0_{\mb{l}})\omega^{s(\mb{g})}_{\mb{h}, \mb{k}, \mb{l}}(^{\overline{\mb{g}}}a)
\eeq

Also, notice that when $\nu=0$, these conditions significantly simplify.
\begin{equation}
    \omega(a_\mb{g},b_\mb{h},c_\mb{k},d_\mb{l})=\omega(\mb{g,h,k,l})\omega_\mb{h,k,l}^{s(\mb{g})}( \g{g}a)\omega_{\mb{k,l}}^{s(\mb{gh})}(\g{gh}a,\g{h}b)\omega_\mb{l}^{s(\mb{ghk})}(\g{ghk}a,\g{hk}b,\g{k}c).
\end{equation}
Eq. \eqref{eqn:4cyc-cond-2} says that, for fixed $\mb{g,h,k}$, $\omega_{\mb{g,h,k}}(a)$ forms a character over $A$, and Eq. \eqref{heiheiha} means that $\omega_{\mb{g,h,k}}(a)$ is a 3-cocycle of $G$ for a fixed $a$. In this case we recover the result of the K\"unneth formula.

\subsection{'t Hooft anomaly formula}
The consistency conditions given in the previous section are not complete. A further condition comes from checking the 4-cocycle conditions for elements $0_\mb{g}, 0_\mb{h}, 0_\mb{k}, 0_\mb{l}, 0_\mb{m}$:
\begin{equation}
    \frac{
    \omega^{s(\mb{g})}(0_\mb{h},0_\mb{k},0_\mb{l}, 0_\mb{m})\omega(0_\mb{g}, [\nu(\mb{h,k})]_\mb{hk},0_\mb{l},0_\mb{m})\omega(0_\mb{g},0_\mb{h},0_\mb{k},[\nu(\mb{l,m})]_{\mb{lm}})
    }{\omega(0_\mb{g}, 0_\mb{h}, 0_\mb{k}, 0_\mb{l})\omega([\nu(\mb{g,h})]_\mb{gh}, 0_\mb{k}, 0_\mb{l},0_\mb{m})\omega(0_\mb{g}, 0_\mb{h}, [\nu(\mb{k,l})]_{\mb{kl}},0_\mb{m})}
    =1
\end{equation}

A straightforward computation yields the following:
\begin{equation}\label{eq: H5 anomaly derivation}
 \frac{\omega^{s(\mb{g})}(\mb{h,k,l,m})\omega(\mb{g,hk,l,m})\omega(\mb{g,h,k,lm})}{\omega(\mb{g,h,k,l})\omega(\mb{gh,k,l,m})\omega(\mb{g,h,kl,m})}\cdot   \mathcal{O}(\mb{g,h,k,l,m})=1
\end{equation}
where the obstruction class $\mathcal{O}$ is defined as
\begin{equation}
	\begin{split}
		\mathcal{O}(\mb{g,h,k,l,m})
		=\omega^{-s(\mb{gh})}_{\mb{k,l,m}}(\g{gh}\nu(\mb{g,h}))\widetilde{\mathcal{O}}(\mb{g,h,k,l,m}),
	\end{split}
	\label{eqn:h5full}
\end{equation}
and
\begin{equation}
\begin{split}
	\widetilde{\mathcal{O}}(\mb{g,h,k,l,m})&=\left(
	\frac{ \omega_{\mb{l,m}}( \ga{ghk}{\nu(\mb{gh,k})},\ga{ghk}{\nu(\mb{g,h})})}{ \omega_{\mb{l,m}}( \ga{ghk}{\nu(\mb{g,hk})},\ga{hk}{\nu(\mb{h,k}))}}\right)^{-s(\mb{ghk})}
	\bigg(
	\frac{\omega_\mb{m}(\g{ghkl}\nu(\mb{gh,kl}),\g{kl}\nu(\mb{k,l}),\g{ghkl}\nu(\mb{g,h}))}{\omega_\mb{m}(\g{ghkl}\nu(\mb{gh,kl}),\g{ghkl}\nu(\mb{g,h}),\g{kl}\nu(\mb{k,l}))}\\
	&\frac{\omega_\mb{m}(\g{ghkl}\nu(\mb{ghk,l}),\g{ghkl}\nu(\mb{g,hk}),\g{hkl}\nu(\mb{h,k}))}{\omega_\mb{m}(\g{ghkl}\nu(\mb{ghk,l}),\g{ghkl}\nu(\mb{gh,k}),\g{ghkl}\nu(\mb{g,h}))}
	\frac{\omega_\mb{m}(\g{ghkl}\nu(\mb{g,hkl}),\g{hkl}\nu(\mb{h,kl}),\g{kl}\nu(\mb{k,l}))}{\omega_\mb{m}(\g{ghkl}\nu(\mb{g,hkl}),\g{hkl}\nu(\mb{hk,l}),\g{hkl}\nu(\mb{h,k}))}
	\bigg)^{-s(\mb{ghkl})}
	\end{split}
	\label{}
\end{equation}
In other words,  for a legitimate $\mc{G}$-SPT phase, $[\mathcal{O}]$ must vanish in $\H^5_s[G, \U]$.

It is a rather nontrivial fact that $\mathcal{O}$ defined in Eq. \eqref{eqn:h5full} is a $5$-cocycle. We present the proof for this result in two cases: (1) when $p=0$, which is always the case if $G$ is unitary and compact/finite, or if $G$ contains anti-unitary elements but the axion angle $\theta=0$. (2) when $G$ commutes with $\U$, \ie $\rho(\mb{g})\equiv s(\mb{g})$ for all $\mb{g}\in G$.

The proof in case (1) involves a direct computation of $\delta_s\mathcal{O}$. In this case,  $\mathcal{O}$ is basically a cup product $\nu\cup n$, as discussed in the main text. Explicitly,
\beq\label{eq: 5-cocycle simpler appendix}
\mathcal{O}(\mb{g,h,k,l,m})
		=\omega^{-s(\mb{gh})}_{\mb{k,l,m}}(\g{gh}\nu(\mb{g,h}))=e^{-is(\mb{gh})^\mb{gh}\nu(\mb{g}, \mb{h})n(\mb{k}, \mb{l},
		\mb{m})}
\eeq

Now we sketch the proof.  We define $\tilde{n}(\mb{g,h,k})=s(\mb{ghk})n(\mb{g,h,k})$. The $\H^3$ obstruction-free condition reads
\begin{equation}
    (\delta_\rho n)(\mb{g,h,k,l})=-\frac{1}{2\pi}2s(\mb{ghk})(\delta [\nu])(\mb{g,h,k})p(\mb{l}).
\end{equation}
Let $\mathcal{O}=e^{i\hat{\mathcal{O}}}$. We split the obstruction class into two parts:
\begin{equation}
    \hat{\mathcal{O}}=\hat{\mc{O}}_1+\hat{\mc{O}}_2.
\end{equation}
Here
\begin{equation}
    \hat{\mc{O}}_1(\mb{g,h,k,l,m})= s(\mb{gh}) \g{gh}\nu(\mb{g,h})n(\mb{k,l,m}),
\end{equation}
and $\hat{O}_2$ contains the rest of the expression, which vanishes if $p=0$.

A direct computation finds
\begin{equation}
    (\delta_s\hat{\mc{O}}_1)(\mb{g,h,k,l,m,n})=-\frac{1}{2\pi}2\rho(\mb{gh})s(\mb{klm}) \nu(\mb{g,h})(\delta_{\rho\cdot s}[\nu])(\mb{k,l,m})p(\mb{n}).
\end{equation}
So this shows that when $p=0$, $\delta_s \hat{O}\equiv 0$. This concludes the proof for case (1). Notice that so far we have not made any further assumptions about $\rho$ and $s$.

Next we compute $\delta_s\hat{O}_2$, now under the assumption that $(\rho\cdot s)(\mb{g})=1$. Explicitly we have:
\begin{equation}
    \hat{\mc{O}}_2(\mb{g,h,k,l,m})= s(\mb{ghkl})\Big[ \nu(\mb{ghk,l})(\delta[\nu])(\mb{g,h,k}) + \nu(\mb{g,hkl})(\delta[\nu])(\mb{h,k,l}) + [\nu(\mb{g,h})][\nu(\mb{k,l})]\Big]p(\mb{m}).
\end{equation}

Through a straightforward but lengthy computation, we obtain
\begin{equation}
    \delta_s \hat{\mc{O}}_2=\frac{1}{2\pi}2s(\mb{ghklm})[\nu(\mb{g,h})]\delta[\nu](\mb{k,l,m})p(\mb{n})
\end{equation}
Therefore,  $\delta_s\mc{O}=\delta_s(\hat{\mc{O}}_1+\hat{\mc{O}}_2)=0$. This concludes the proof for case (2). So $\mc{O}$ is a 5-cocycle if $p=0$ or if $G$ commutes with $\U$. 

In summary, in order for a given symmetry fractionalization pattern characterized by the triple $(\rho, \nu, n)$ to be anomaly-free, both Eq. \eqref{eqn:eqforn} and Eq. \eqref{eq: H5 anomaly derivation} must hold. We believe these two equations also form a sufficient condition for the the triple $(\rho, \nu, n)$ to be anomaly-free.

\section{Hall conductivity of a \spd{2} U($N$) symmetric invertible states} \label{app: Hall}

In this appendix we discuss the constraint from the U($N$) symmetry on the Hall conductivity of a \spd{2} bosonic invertible states (at the end we also briefly discuss the similar constraint on fermionic invertible states). We will see that if the boson is in the (bi-)fundamental represention of U($N$) (\ie all charge-1 bosons also carry the fundamental representation of SU($N$)), then the minimal nonzero Hall conductivity of an invertible state is $N$ ($2N$) in units of $e^2/h$, if $N$ is even (odd). The simplest way to see this is to consider gauging the $\U$ symmetry and examine the $2\pi$ instanton operator. This instanton operator should be (1) bosonic and (2) carry a linear representation of PSU($N$). Condition (1) means that $\sigma_{xy}$ is even, and condition (2) means that $\sigma_{xy}$ is an integer multiple of $N$. Therefore, the minimal nonzero $\sigma_{xy}$ is $N$ ($2N$), if $N$ is even (odd). This result of course agrees with Refs. \cite{Liu2012, Senthil2012}, where the special cases with $N=1,2$ have been discussed.

To show this result more formally, we generalize the argument in Ref. \cite{Seiberg2016}, which was applied to the special cases with $N=1,2$ therein. The Hall conductivity can be determined by the response theory of this bosonic invertible state to an external U($N$) gauge field, $\mb{a}=a+\tilde a\mb{1}$ with $a$ an SU($N$) gauge field and $\tilde a$ the $\U$ gauge field. The generic (topological) response can be captured by the Chern-Simons Lagrangian:
\beq
\begin{split}
\mc{L}
&=\frac{k_1}{4\pi}\tilde a d\tilde a+\frac{k_2}{4\pi}\Tr\left(ada+\frac{2}{3}a^3\right)\\
&=\frac{k_2}{4\pi}\Tr\left(\mb{a}d\mb{a}+\frac{2}{3}\mb{a}^3\right)+\frac{k_1-Nk_2}{4\pi N^2}\left(\Tr\mb{a}\right)d\left(\Tr\mb{a}\right)
\end{split}
\eeq
Notice $\sigma_{xy}=k_1$ in units of $e^2/h$. Below we determine the possible values of $k_1$.

In order for this Lagrangian to describe a valid response of a bosonic invertible state, we can consider the case where $\mb{a}={\rm diag}(a_{11}, 0, 0,\cdots, 0)$. Then the above Lagrangian becomes
\beq
\mc{L}=\frac{k_1+N(N-1)k_2}{4\pi N^2}a_{11}da_{11}.
\eeq
For this to be a valid bosonic response, there must exist an integer $m$, such that 
\beq \label{eq: Hall constraint}
k_1+N(N-1)k_2=2mN^2
\eeq
Clearly $k_1$ is a multiple of $N$, so we can write it as $k_1=N\cdot n$, with $n$ an integer. Now our goal becomes to find the possible values of $n$, which satisfies
\beq
n+(N-1)k_2=2mN
\eeq
The right hand side is even, so must be the left hand side. 

If $N$ is odd, then $n$ must be even for the left hand side to be even. The smallest nonzero even number is $2$, and $n=2$ can be achieved by having $k_2=2$ and $m=1$. So if $N$ is odd, the minimal nonzero Hall conductivity is $\sigma_{xy}=2N$. If $N$ is even, then it is possible to achieve $n=1$ by having $k_2=N+1$ and $m=N/2$. So the minimal nonzero Hall conductivity is $\sigma_{xy}=N$ if $N$ is even.

Suppose one wants to have a bosonic system with a nonzero $\sigma_{xy}$ smaller than $N$ ($2N$) for even (odd) $N$, besides making the bosons form a fractional quantum Hall state, one can also consider making the bosons living on the boundary of a \spd{3} system with a bulk $\theta$-term for the $\U$ gauge field:
\beq
\frac{\theta}{8\pi^2}\tilde F\wedge \tilde F
\eeq
with $\theta=2\pi\sigma_{xy}$ and $\tilde F$ the $\U$ gauge field strength that is extended into the \spd{3} bulk. In the absence of any other global symmetries, this \spd{3} bulk is a generic invertible state (\ie it can be smoothly connected to a product state without encountering a phase transition). Nevertheless, its boundary is an invertible state that cannot be realized in purely \spd{2}. Such a boundary state is also referred to as an anomalous invertible state \cite{Wang2018}.

For completeness, we also discuss the minimal Hall conductivity for a U($N$) symmetric fermionic invertible state. In this case, the constraint equation Eq. (\ref{eq: Hall constraint}) is replaced by
\beq
k_1+N(N-1)k_2=mN^2
\eeq
Here $k_1=N$ can always be achieved by having $k_2=m=1$. That is to say, the minimal nonzero Hall conductivity for a U($N$) symmetric fermionic invertible state is $\sigma_{xy}=N$ for any $N$. Physically, this state can be realized by a U($N$) symmetric version of the Haldane model (\ie in the field-theoretic terminology, a pair of gapped Dirac fermions in the (bi-)fundamental representation of U$(N)$ symmetry) \cite{Haldane1988}. Just as the bosonic case, at the expense of putting it on the surface of a \spd{3} bulk with a $\theta$-term, here one can also have anomalous fermionic invertible states where $\sigma_{xy}$ is smaller than $N$.

\section{Derivation of $\Z_2\times \Z_2^\mathsf{T}$ anomaly} 
\label{app: Z2 x Z2T}

Here we present details of the classification of anomalous U(1) spin liquids enriched by $\Z_2\times \Z_2^\mathsf{T}$ symmetry. 

\subsection{K\"unneth formula and slant product}
It is known that $\H^5[\Z_2 \times \Z_2^\mathsf{T}, \U]=\Z_2^3$, which means that there are three root bosonic SPT states in \spd{4} with $\Z_2\times \Z_2^\mathsf{T}$ symmetry. This can be understood by the  K\"unneth formula for group cohomology: 
\begin{equation}
	\begin{split}
	\H^5[\Z_2\times \Z_2^\mathsf{T}, \U]
	&=\H^5[\Z_2, \H^0[\Z_2^\mathsf{T}, \U]]  \oplus \H^3[\Z_2, \H^2[\Z_2^\mathsf{T}, \U]] \nonumber  \oplus \H^1[\Z_2, \H^4[\Z_2^\mathsf{T}, \U]] \nonumber \\
 &= \Z_2 \oplus \Z_2 \oplus \Z_2
	\end{split}
	\label{}
\end{equation}

In the following we write down representative cocycles for $\mc{O}(\mb{g, h, k, l, m})$. We will denote the group $\Z_2\times\Z_2^\mathsf{T}$ additively, \ie $\mb{g}=(g_1,g_2)$ where $g_{1,2}=\{0,1\}$. $Z\equiv (1,0)$ generates $\Z_2$ and $\mathsf{T}\equiv (0,1)$ generates $\Z_2^\mathsf{T}$.

The first $\Z_2$ from $\H^5[\Z_2,\H^0[\Z_2^\mathsf{T}, \U]]$ corresponds to a root state protected solely by the $\Z_2$ subgroup of the symmetry group.   A representative cocycle for this class is
\begin{equation}
	\mathcal{O}_1(\mb{g, h, k, l, m})=(-1)^{g_1h_1k_1l_1m_1}.
	\label{}
\end{equation}
This expression is equivalent to the topological partition function $\exp\left(i\pi\int a^5\right)$, where $a$ is a 1-cocycle representing the $\Z_2$ gauge field. This state is precisely the one discussed in Sec. \ref{subsec: Z2}, whose boundary can be $(\E_{bZ}\M_{bZ})_-$.

The second $\Z_2$, from $\H^1[\Z_2,\H^4[\Z_2^\mathsf{T}, \U]]$, is generated by the following cocycle:
\begin{equation}
	\mathcal{O}_2(\mb{g,h,k,l,m})=(-1)^{g_1h_2k_2l_2m_2}.
	\label{}
\end{equation}
This expression is equivalent to the partition function $\exp\left(i\pi\int a\cup w_1^4\right)$, where $w_1$ is the first Stieffel-Whitney class of the tangent bundle.

Similarly, the last $\Z_2$ factor is generated by
\begin{equation}
	\mathcal{O}_3(\mb{g,h,k,l,m})=(-1)^{g_1h_1k_1l_2m_2}.
	\label{eqn:O3-1}
\end{equation}
The corresponding partition function is $\exp\left(i\pi\int a^3\cup w_1^2\right)$. To simplify our calculation, notice that $[a^3\cup w_1^2]=[w_1^2\cup a^3]$ ($[\cdot]$ means the cohomology class). So an equivalent expression for $\mathcal{O}_3$ is given by
\begin{equation}
	\mathcal{O}_3(\mb{g,h,k,l,m})=(-1)^{g_2h_2k_1l_1m_1}.
	\label{eqn:O3-2}
\end{equation}

We now prove that $[\mathcal{O}_{1,2,3}]$ are indeed distinct cohomology classes. Clearly, $[\mc{O}_1]$ is different from $[\mc{O}_{2,3}]$, because the former is nontrivial even in the absence of the $\Z_2^\TT$ symmetry, while the latter requires the $\Z_2^\TT$ to be nontrivial.

Next, we consider $\mc{O}_2$. Using 0-slant product, we find
\begin{equation}
	(i_Z\mathcal{O}_2)(\mb{h,k,l,m}) = (-1)^{h_2k_2l_2m_2}.
	\label{}
\end{equation} 
Thus the restriction $\left.i_Z\mathcal{O}_2\right|_{\Z_2^\TT}$ belongs to the nontrivial cohomology class in $\H^4[\Z_2^\TT, \U]$ (\ie $(i_Z\mathcal{O}_2)(\mathsf{T,T,T,T})=-1$). Physically, this means that the $\Z_2$ domain wall of the corresponding \spd{4} SPT state hosts a \spd{3} time reversal SPT state $eTmT$. In passing, we note that decorating another \spd{3} time reversal SPT state $efmf$ onto the $\Z_2$ domain wall results in a \spd{4} $\Z_2$ SPT that is beyond group cohomology \cite{Wen2014}.

Now we consider $\mathcal{O}_3$. Using the representation of $\mathcal{O}_3$ in Eq. \eqref{eqn:O3-2}, one finds $i_Z\mathcal{O}_3=1$. Thus $[\mathcal{O}_3]\neq [\mathcal{O}_2]$. We still need to show that $[\mathcal{O}_3]$ is nontrivial. To this end, let us consider $2$-slant product $i_{\mathsf{T,T}}\mathcal{O}_3$. It is easy to see that $(i_\mathsf{T}\mathcal{O}_3)(\mb{h,k,l,m})=1$, so $i_{\mathsf{T,T}}\mathcal{O}_3$ is a $3$-cocycle of $\Z_2\times\Z_2^\mathsf{T}$. We find
\begin{equation}
	(i_{\mathsf{T,T}}\mathcal{O}_3)(\mb{k,l,m})=(-1)^{k_1l_1m_1},
	\label{}
\end{equation}
which is a nontrivial $3$-cocycle of the $\Z_2$ subgroup. Thus $\mathcal{O}_3$ indeed belongs to a nontrivial cohomology class.

A general $5$-cocycle $\mathcal{O}$ can be decomposed as
\begin{equation}
	[\mathcal{O}]=[\mathcal{O}_1]^{\frac{1-r_1}{2}}\cdot [\mathcal{O}_2]^{\frac{1-r_2}{2}}\cdot [\mathcal{O}_3]^{\frac{1-r_3}{2}}, r_{1,2,3}=\pm 1.
	\label{eq: Z2Z2T decomposition}
\end{equation}
Here
\begin{equation}
	\begin{split}
	r_1&=\mathcal{O}(Z,Z,Z,Z,Z),\\
	r_2&=(i_Z\mathcal{O})(\mathsf{T,T,T,T}),\\
	r_3&=(i_{\mathsf{T,T}}\mathcal{O})(Z,Z,Z).
	\end{split}
	\label{}
\end{equation}
It can be readily checked that all three are invariants of cohomology classes.

\subsection{Fractionalization classes}

In this section we classify projective representations of $\Z_2\times\Z_2^\TT$.
Note that the projective representations of eletric charge are classified by the second cohomology class $\H_{\rho\cdot s}^2[G, \U]$, while those of the magnetic charge are classified by $\H_{\rho}^2[G, \U]$. The subscript $\rho\cdot s$ or $\rho$ denote the group action on the $\U$ coefficient. Below we would discuss different cases in details. 
\begin{enumerate}
	\item $\rho(Z)=\mathds{1}$: projective representions of an electric charge is classified by $\H^2_{s}[\Z_2\times \Z_2^\mathsf{T}, \U]=\Z_2^2$, whose generators  can be represented by 
		\begin{equation}
			\nu(\mb{g, h})=(-1)^{g_1h_1} \text{ or } (-1)^{g_2h_2}.
			\label{}
		\end{equation}
		For dual magnetic charges, the projective representations are classified by $\H^2[\Z_2\times\Z_2, \U]=\Z_2$ (namely, the coefficient $\U$ is a module with trivial action). The genertor is represented by 
		\begin{equation}
			\omega_\M(\mb{g, h})=(-1)^{g_1h_2}.
			\label{}
		\end{equation}

	\item $\rho(Z)=-\mathds{1}$:  the projective representions of electric charges  are classified by $\H^2[\Z_2^\mathsf{T}\times \Z_2^\mathsf{T}, \U]=\Z_2^2$ whose generators can be represented 
	\begin{equation}
	    \nu(\mb{g,h})=(-1)^{g_1h_1} \text{ or }(-1)^{g_2h_2}.
	\end{equation}
	 While for magnetic monopoles, the projective representations are classified by $\H^2[\Z_2^\mathsf{T}\times\Z_2, \U]=\Z_2^2$, generated by
	 \begin{equation}
	    \omega_\M(\mb{g,h})=(-1)^{g_1h_1} \text{ or }(-1)^{g_2h_2}.
	\end{equation}
\end{enumerate}

Using these explicit parametrizations of $\nu$ and $\omega_\M$, the expression for $\mc{O}$ in Eq. \eqref{eq: 5-cocycle simpler appendix}, the definitions of the slant products, and Eq. \eqref{eq: Z2Z2T decomposition}, it is straightforward to obtain the results in Table \ref{tab:z2z2t}.

\section{Derivation of LSM anomaly}\label{app: LSM anomaly}

We present more details of the discussion on the LSM anomaly. We will in fact consider a more general unitary on-site symmetry group $H$, assuming that $H$ acts trivially on the charge types. Now the full symmetry group is $G=H\times\Z^3$. We further assume that $\H^2_\rho[G, \U]=\H^2_\rho[H, \U]\times\H^2_\rho[\Z^3, \U]$, which allows us to write a generic 2-cocycle $\nu\in\H_\rho^2[G, \U]$ in the form of $\nu=(\nu_1, \nu_2)$, where $\nu_1\in\H^2_\rho[H, \U]$ and $\nu_2\in\H^2_\rho[\Z^3, \U]$, \ie the symmetry fractionalization pattern of $G$ can be decomposed into that of $H$ and that of $\Z^3$.

Parallel to the discussion in Sec. \ref{sec:LSManomaly}, we will only consider the case where translation acts as charge conjugation. In this case, $\H^2_\rho[\Z^3, \U]=\Z_2$, which means there is only one nontrivial translation fractionalization pattern. The 2-cocycle of this nontrivial translation fractionalization pattern can be written as
\beq
\eta(\mb{a}, \mb{b})=(-1)^{b_x(a_y+a_z)+b_y(a_x+a_y)}
\eeq
where $\mb{a}=T_x^{a_x}T_y^{a_y}T_z^{a_z}$, with $T_{x,y,z}$ the generator of translation along the $x,y,z$ directions, and $a_{x,y,z}\in\Z$ (similar for $\mb{b}$). It is straightforward to check that this is indeed a 2-cocycle, and the invariant is nontrivial
\beq
\frac{\eta(T_x,T_y)}{\eta(T_y, T_x)}\frac{\eta(T_y,T_z)}{\eta(T_z, T_y)}\frac{\eta(T_z,T_x)}{\eta(T_x, T_z)}=-1
\eeq

Below we analyze whether a symmetry-enriched $\U$ QSL satisfies the LSM constriant, \ie we will check what projective representation of $H$ each unit cell carries. To do so, we can calculate 
\beq
\nu_{xyz}\equiv i_{T_z}i_{T_y}i_{T_x}\mc{O}
\eeq
where $\mc{O}$ is the obstruction 5-cocycle of this symmetric $\U$ gauge theory. Then $\left.\nu_{xyz}\right|_H$,  the restriction of $\nu_{xyz}$ to $H$, represents the representation of $H$ in each unit cell.

The following observation will simplify the calculation of $\left.\nu_{xyz}\right|_H$. With the symmetry $G$, a generic $\U$ gauge theory can be written as $(\E_{b(\nu_1,\nu_2)}\M_{b(\mu_1,\mu_2)})_-$, where the $-$ outside the bracket reminds us that the translations act as charge conjugation, and the subscript $b(\nu_1, \nu_2)$ means that this excitation is a boson with symmetry fractionalization pattern $\nu=(\nu_1, \nu_2)$. The anomaly of $(\E_{b(\nu_1,\nu_2)}\M_{b(\mu_1,\mu_2)})_-$ can be decomposed into the anomalies of some other states:
\beq
\begin{split}
&(\E_{b(\nu_1,\nu_2)}\M_{b(\mu_1,\mu_2)})_-\\
=&(\E_{b(\nu_1,\nu_2)}\M_{b(0,\mu_2)})_-\oplus(\E_{b(\nu_1,\nu_2)}\M_{b(\mu_1,0)})_-\\
=&(\E_{b(\nu_1,0)}\M_{b(0,\mu_2)})_-\oplus(\E_{b(0,\nu_2)}\M_{b(0,\mu_2)})_-\oplus (\E_{b(\nu_1,0)}\M_{b(\mu_1,0)})_-\oplus(\E_{b(0,\nu_2)}\M_{b(\mu_1,0)})_-
\end{split}
\eeq
where $A\oplus B$ means adding the anomalies of $A$ and $B$, or more physically, stacking $A$ and $B$ and switching on the hybridization of their certain excitations. The above decomposition means that, in order to obtain $\left.\nu_{xyz}\right|_H$ for $(\E_{b(\nu_1,\nu_2)}\M_{b(\mu_1,\mu_2)})_-$, we just need to obtain the $\left.\nu_{xyz}\right|_H$'s of the four simpler states and then add them together.

It is quite straightforward to see that $\left.\nu_{xyz}\right|_H=0$ for both $(\E_{b(0,\nu_2)}\M_{b(0,\mu_2)})_-$ and $(\E_{b(\nu_1,0)}\M_{b(\mu_1,0)})_-$. So our task becomes to determine $\left.\nu_{xyz}\right|_H$ for $(\E_{b(\nu_1,0)}\M_{b(0,\mu_2)})_-$ and $(\E_{b(0,\nu_2)}\M_{b(\mu_1,0)})_-$. Also notice the state $(\E_{b(0,\nu_2)}\M_{b(\mu_1,0)})_-$ is really identical to the state $(\E_{b(-\mu_1,0)}\M_{b(0,\nu_2)})_-$, because the latter is related to the former by a relabelling: $\E\rightarrow\M^\dag$ and $\M\rightarrow\E$. Now $(\E_{b(\nu_1,0)}\M_{b(0,\mu_2)})_-$ and $(\E_{b(-\mu_1,0)}\M_{b(0,\nu_2)})_-$ have the same form of symmetry fractionalization pattern, so it suffices to calculate $\left.\nu_{xyz}\right|_H$ for one of them. We will calculate $\left.\nu_{xyz}\right|_H$ for $(\E_{b(\nu_1,0)}\M_{b(0,\mu_2)})_-$.

The obstruction 5-cocycle $\mc{O}$ for $(\E_{b(\nu_1,0)}\M_{b(0,\mu_2)})_-$ is given by Eq. \eqref{eq: 5-cocycle simpler appendix}. If any of $\nu_1$ and $\mu_2$ is trivial, $\mc{O}=1$ is trivial, and $\left.\nu_{xyz}\right|_H=0$. On the other hand, if $\nu_1$ is nontrivial and $\mu_2=\eta$, we can get the 3-cocycle representation of $\mu_2$:
\beq
n_2(\mb{a,b,c})=(\delta \hat\mu_2)(\mb{a,b,c}) = \frac{1}{2}(1-(-1)^{a_x+a_y+a_z})\left[c_x(b_y+b_z)+c_y(b_x+b_y)\right]
\eeq
where $\mu_2(\mb{a}, \mb{b})=\eta(\mb{a}, \mb{b})=e^{2\pi i\hat\mu_2(\mb{a}, \mb{b})}$. Plugging the nontivial $\nu_1$ and $n_2$ into Eq. \eqref{eq: 5-cocycle simpler appendix} yields $\mc{O}$, and a straightforward but lengthy calculation yields that $\left.\nu_{xyz}\right|_H=-\nu_1$. Taken these cases together, we can write, for $(\E_{b(\nu_1,0)}\M_{b(0,\mu_2)})_-$
\beq
\left.\nu_{xyz}\right|_H=\frac{\lambda_{\mu_2}-1}{2}\cdot\nu_1
\eeq
where $\lambda_{\chi}$ is the cohomological invariant of a translation fractionalization class $\chi$:
\beq
\lambda_{\chi}\equiv\frac{\chi(T_x,T_y)}{\chi(T_y, T_x)}\frac{\chi(T_y,T_z)}{\chi(T_z, T_y)}\frac{\chi(T_z,T_x)}{\chi(T_x, T_z)}
\eeq

The above results also indicate that, for $(\E_{b(-\mu_1,0)}\M_{b(0,\nu_2)})_-$, we have
\beq
\left.\nu_{xyz}\right|_H=\frac{1-\lambda_{\nu_2}}{2}\cdot\mu_1
\eeq
Taking all the above results together yields $\left.\nu_{xyz}\right|_H$ for $(\E_{b(\nu_1,\nu_2)}\M_{b(\mu_1,\mu_2)})_-$:
\beq \label{eq: LSM matching equation}
\left.\nu_{xyz}\right|_H=\frac{\lambda_{\mu_2}-1}{2}\cdot\nu_1+\frac{1-\lambda_{\nu_2}}{2}\cdot\mu_1
\eeq

This result means that, suppose each unit cell has a representation of $H$ specified by the factor set $\left.\nu_{xyz}\right|_H$, the LSM constraint enforces that the symmetry-enriched $\U$ gauge theory can be $(\E_{b(\nu_1,\nu_2)}\M_{b(\mu_1,\mu_2)})_-$, if Eq. \eqref{eq: LSM matching equation} is satisfied. Notice that the state $(\E_{b(\nu_1,\nu_2)}\M_{b(\mu_1,\mu_2)})_-$ may suffer from other anomalies that is not of the LSM-type, and these anomalies need to be examined separately.

Now let us apply this result to the case with $H=\mathrm{SO}(3)$. By enumerating all possible fractionalization patterns, we have the following list of 7 possible symmetric $\U$ QSLs: $(\E_b\M_b)_-$, $(\E_{b\half}\M_b)_-$, $(\E_{b\trn}\M_b)_-$, $(\E_{b\half\trn}\M_b)_-$, $(\E_{b\half}\M_{b\trn})_-$, $(\E_{b\trn}\M_{b\trn})_-$ and $(\E_{b\half\trn}\M_{b\trn})_-$. Notice we have already removed states where both $\E$ and $\M$ are spin-1/2 bosons, because they are anomalous (see Sec. \ref{sec:so3} and Ref. \cite{Zou2017}). Our analysis of the LSM anomaly indicates that only $(\E_{b\half}\M_{b\trn})_-$ and $(\E_{b\half\trn}\M_{b\trn})_-$ can possibly be realized on a lattice with an odd number of spin-1/2's per unit cell, and the other 5 states can only be realized on a lattice with an even number of spin-1/2's per unit cell. The explicit parton constructions in Appendix \ref{app: parton} imply that all these states can indeed be realized in their corresponding types of lattice systems, \ie there is no more anomaly.

\section{Parton constructions of $(\E_{b\half}\M_{b\trn})_-$ and $(\E_{f\half}\M_{b\trn})_-$} \label{app: parton}

In this appendix we carry out explicit parton constructions of $(\E_{b\half}\M_{b\trn})_-$ and $(\E_{f\half}\M_{b\trn})_-$, where the subscript $f\half$ indicates a fermionic spin-1/2 excitation. These states are $\U$ QSLs with SO(3) spin rotational symmetry and translation symmetry. Notice in the present symmetry setting the state $(\E_{f\half}\M_{b\trn})_-$ is equivalent to $(\E_{b\half\trn}\M_{b\trn})_-$, so it is sufficient to realize the former to show that the latter is also realizable. From our consideration based on the quantum anomalies, they can emerge only if the microscopic lattice system has a half-odd-integer spin in each unit cell. The converse is also true, \ie if a $\U$ QSL emerges from a lattice system with a half-odd-integer spin per unit cell and the system preserves the SO(3) and translation symmetries, the symmetry fractionalization patterns of these symmetries must be that the resulting symmetry-enriched $\U$ QSL is either $(\E_{b\half}\M_{b\trn})_-$ or $(\E_{f\half}\M_{b\trn})_-$. The purpose of this section is to show that these two states can indeed emerge.

\subsection{Parton construction of $(\E_{b\half}\M_{b\trn})_-$}

We begin with $(\E_{b\half}\M_{b\trn})_-$, and we will take the electric charge $\E$ as the parton. We will consider a cubic lattice with a spin-1/2 on each site, and we will write the spin operators in terms of Schwinger bosons:
\beq
\vec S_i=\frac{1}{2}b_i^\dag\vec\sigma b_i
\eeq
where $i$ labels the site, $\vec\sigma$ is the standard Pauli matrices, and $b_i=(b_{i1}, b_{i2})^T$ labels a canonical boson that transforms in the fundamental representation of SU(2). To faithfully represent the original spin system, a gauge constraint needs to be imposed:
\beq \label{eq: gauge constraint bosons}
b_i^\dag b_i=1
\eeq
These Schwinger bosons will be regarded as the electric charges. The task of having a parton construction of $\E_{b\half}\M_{b\trn}$ becomes to put these Schwinger bosons into a gapped state that preserve the symmetries and gauge constraint.

To do so, we follow Refs. \cite{Read1989, Read1990}. First, we divide the cubic lattice into the A and B sublattices in the usual way, and on sublattice B we introduce operators $\bar b_i$:
\beq
\bar b_{i1}\equiv b_{i2},
\quad
\bar b_{i2}\equiv -b_{i1}
\eeq
In the following, operators on sublattice A will still be expressed in terms of $b$'s, while operators on sublattice B will always be expressed in terms of $\bar b$'s. Notice now the gauge constraint on sublattice A is still given by Eq. (\ref{eq: gauge constraint bosons}), and on sublattice B it becomes
\beq \label{eq: gauge constraint bosons sublattice B}
\bar b_i^\dag\bar b_i=1
\eeq

Next, consider putting the Schwinger bosons into a state described by the following (mean-field) Hamiltonian:
\beq \label{eq: mean field boson}
H=\lambda\sum_{i\in A}(b^\dag_ib_i-1)+\lambda\sum_{i\in B}(\bar b_i^\dag \bar b_i-1)-\sum_{i\in A,\hat\eta}(Qb^T_i\bar b_{i+\hat\eta}+\hc)
\eeq
where $\hat\eta$ labels the vectors connecting a pair of adjacent sites that are on sublattices A and B, respectively. The parameter $\lambda$ is responsible for imposing the gauge constraint Eq. (\ref{eq: gauge constraint bosons}) at the level of expectation value, and the parameter $Q$ is a coupling strength that is determined by interactions between the Schwinger bosons and can be tuned by tuning the microscopic parameters of the original lattice system.

Now, let us examine the symmetries of this model. It is straightforward to see that the SO(3) symmetry is preserved, and both $b$ and $\bar b$ carry spin-1/2. The translation symmetry along $\hat\eta$ is also preserved, and it acts as
\beq
b_i\rightarrow \bar b_{i+\hat\eta},
\quad
\bar b_i\rightarrow b_{i+\hat\eta}
\eeq
Although there is a term $b^T_i\bar b_{i+\hat\eta}$ in the Hamiltonian, due to the bipartite nature of the system there is still a $\U$ symmetry where $b$ and $\bar b$ carry opposite charges:
\beq
b_i\rightarrow e^{i\theta}b_i,
\quad
\bar b_i\rightarrow e^{-i\theta}\bar b_i
\eeq
From here we also see that the action of translation symmetry has the effect of charge-conjugation on the $\U$ charges, but it belongs to the trivial class in $\H^2_{\rho}[\Z^3; \U]$ with $\rho=-1$ on generators of translations. From these we conclude that the Schwinger bosons here have the same projective symmetry properties as that of the electric charges in $(\E_{b\half}\M_{b\trn})_-$.

Finally, we only need to demonstrate that the Hamiltonian (\ref{eq: gauge constraint bosons}) can be gapped while maintaining the gauge constraints (\ref{eq: gauge constraint bosons}) and (\ref{eq: gauge constraint bosons sublattice B}) at the level of expectation values. To do so, we first write down the momentum-space representation of the Hamiltonian (\ref{eq: mean field boson}):
\beq \label{eq: mean field boson momentum space}
H=\lambda\sum_{\vec k}(b^\dag_{\vec k}b_{\vec k}+\bar b_{\vec k}^\dag b_{\vec k}-2)-Q\sum_{\vec k}(b^T_{\vec k}\bar b_{-\vec k}\gamma_{\vec k}+\hc)
\eeq
with $\gamma_{\vec k}=\sum_{\hat\eta}\exp(-i\vec k\cdot\hat\eta)$ (which for a cubic lattice is $\gamma_{\vec k}=2(\cos k_x+\cos k_y+\cos k_z)$ if the lattice constants are all taken to be unity). Next, we apply the standard Bogoliubov transformation:
\beq
b(\vec k)=\eta_{\vec k}\cosh\theta_{\vec k}-\eta_{-\vec k}^*\sinh\theta_{\vec k},
\quad
\bar b(\vec k)=\eta_{-\vec k}\cosh\theta_{\vec k}-\eta_{\vec k}^*\sinh\theta_{\vec k}
\eeq
where $\eta_{\vec k}$'s are canonical bosons with commutators, \eg $[\eta_{\vec k\alpha}, \eta_{\vec q\beta}^*]=\delta_{\vec k\vec q}\delta_{\alpha\beta}$, where $\alpha$ and $\beta$ are spin indices. In terms of $\eta$'s, the Hamiltonian (\ref{eq: mean field boson momentum space}) becomes (up to constant terms)
\beq
\begin{split}
H=
&\sum_{\vec k}\eta_{\vec k}^\dag\eta_{\vec k}[2\lambda(\cosh^2\theta_{\vec k}+\sinh^2\theta_{\vec k})+4Q\gamma_{\vec k}\sinh\theta_{\vec k}\cosh\theta_{\vec k}]\\
+&\sum_{\vec k}
\left\{
\eta_{\vec k}\eta_{-\vec k}[-2\lambda\sinh\theta_{\vec k}\cosh\theta_{\vec k}-Q\gamma_{\vec k}(\cosh^2\theta_{\vec k}+\sinh^2\theta_{\vec k})]+\hc
\right\}
\end{split}
\eeq
By requiring that
\beq
\tanh 2\theta_{\vec k}=-\frac{Q\gamma_{\vec k}}{\lambda}
\eeq
the above Hamiltonian simplifies to
\beq
H=\sum_{\vec k}E_{\vec k}\eta_{\vec k}^\dag\eta_{\vec k}
\eeq
with the dispersion of the Bogoliubov quasiparticles given by
\beq
E_{\vec k}=2\sqrt{\lambda^2-Q^2\gamma_{\vec k}^2}
\eeq
The condition of being gapped is 
\beq \label{eq: gap boson}
\Delta\equiv 2\sqrt{\lambda^2-Q^2\gamma_0^2}>0
\eeq
and the gauge constraints become
\beq \label{eq: gauge constraint bosons final}
\sum_{\vec k}\sinh^2\theta_{\vec k}=\frac{1}{2}N_{{\rm U.C.}}
\eeq
where $N_{{\rm U.C.}}$ is the number of unit cells in the system, and the factor of $1/2$ is due to the two spins. By tuning $\gamma$ and $Q$, both (\ref{eq: gap boson}) and (\ref{eq: gauge constraint bosons final}) can be satisfied.

Therefore, the Schwinger bosons with the desired symmetry properties can be put into a gapped state. When the U(1) symmetry is gauged, the system becomes a U(1) QSL. Ref. \cite{Motrunich2004} shows that the translation symmetry action on the monopole (which is a boson) of this U(1) QSL belongs to the nontrivial class in $\H^2_{\rho}(\Z^3, \U]$ with $\rho=-1$. We also know when both the electric charge and magnetic monopoles are bosons and the charge carries spin-1/2 under the SO(3) symmetry, the magnetic monopole must carry integer spin. Therefore, the gauged symmetry-enriched U(1) QSL is precisely $(\E_{b\half}\M_{b\trn})_-$.

\subsection{Parton construction of $(\E_{f\half}\M_{b\trn})_-$}

Next we turn to $(\E_{f\half}\M_{b\trn})_-$, and we will again take the electric charge $\E$ as the parton. The construction is very similar to that of $(\E_{b\half}\M_{b\trn})_-$, but now we will consider Abrikosov fermions on a cubic lattice with a spin-1/2 per site. More precisely, we write the spin operators in terms of Abrikosov fermions:
\beq
\vec S_i=\frac{1}{2}f^\dag_i\vec\sigma f_i
\eeq
with $f_i=(f_{i1}, f_{i2})^T$ fermionic operators transforming in the fundamental representation of $SU(2)$. Again, a gauge constraint needs to be imposed in order to represent the system faithfully:
\beq \label{eq: gauge constraint fermions}
f^\dag_if_i=1
\eeq
These Abrikosov fermions will be regarded as the electric charges. To construct $\E_{f\half}\M_{b\trn}$, we need to show that these fermions can be put into a gapped state with all the relevant symmetries and gauge constraint.

Again, we will express the operators on sublattice B with a new set of operators:
\beq
\bar f_{i1}=f_{i2},
\quad
\bar f_{i2}=-f_{i1}
\eeq
In terms of $\bar f$'s, the gauge constraint on sublattice $B$ is given by
\beq \label{eq: gauge constraint fermions sublattice B}
\bar f^\dag_i\bar f_i=1
\eeq
while on sublattice A it is still given by (\ref{eq: gauge constraint fermions}).

Now consider putting the Abrikosov fermions into a state described by the following Hamiltonian:
\beq \label{eq: mean field fermion}
H=\lambda\sum_{i\in A}(f^\dag_if_i-1)+\lambda\sum_{i\in B}(\bar f_i^\dag \bar f_i-1)-\sum_{i\in A,\hat\eta}(Qf^T_i\bar f_{i+\hat\eta}+\hc)
\eeq
The symmetries of this Hamiltonian include SU(2) where the fermions carry spin-1/2, translation that acts as
\beq
f_i\rightarrow\bar f_{i+\hat\eta},
\quad
\bar f_i\rightarrow -f_{i+\hat\eta}
\eeq
and U(1) that acts as:
\beq
f_i\rightarrow e^{i\theta}f_i,
\quad
\bar f_i\rightarrow -e^{-i\theta}\bar f_i
\eeq
Again, we can see the fermionic partons carry the same projective symmetry quantum numbers as that of the electric charge of $\E_{f\half}\M_{b\trn}$. Below we only need to show that these fermions can be in a gapped state with the symmetries and gauge constraints preserved.

The momentum-space representation of the Hamiltonian (\ref{eq: mean field fermion}) is
\beq \label{eq: mean field fermion momentum space}
H=\lambda\sum_{\vec k}(f^\dag_{\vec k}f_{\vec k}+\bar f_{\vec k}^\dag f_{\vec k}-2)-Q\sum_{\vec k}(f^T_{\vec k}\bar f_{-\vec k}\gamma_{\vec k}+\hc)
\eeq
This Hamiltonian can again be solved by the standard Bogoliubov transformation:
\beq \label{eq: mean field fermion momentum space}
f(\vec k)=d_{\vec k}^{(1)}\cos\theta_{\vec k}-d_{\vec k}^{(2)}\sin\theta_{\vec k},
\quad
\bar f(-\vec k)^*=d_{\vec k}^{(1)}\sin\theta_{\vec k}+d_{\vec k}^{(2)}\cos\theta_{\vec k}
\eeq
with $\cos 2\theta_{\vec k}=\frac{\lambda}{\sqrt{\lambda^2+Q^2\gamma_{\vec k}^2}}$. In terms of the $d$'s, the Hamiltonian (\ref{eq: mean field fermion momentum space}) becomes
\beq
H=\sum_{\vec k}E_{\vec k}(d_{\vec k}^{(1)\dag} d_{\vec k}^{(1)}-d_{\vec k}^{(2)\dag} d_{\vec k}^{(2)})
\eeq
with the dispersion of the Bogoliubov quasiparticles given by
\beq
E_{\vec k}=\sqrt{\lambda^2+Q^2\gamma_{\vec k}^2}
\eeq
which is always gapped. The gauge constraint now becomes
\beq
\sum_{\vec k}\sin^2\theta_{\vec k}=\frac{1}{2}N_{{\rm U.C.}}
\eeq
which can be satisfied by tuning $\lambda$ and $Q$. 

So we conclude that these Abrikosov fermions can be gapped while preserving the relevant symmetries and gauge constraints. Notice when both $\lambda$ and $Q$ are nonzero, the gauge structure of this parton mean field is $\U$ \cite{Wen2004Book}. When the $\U$ symmetry is gauged, we again get a $\U$ QSL. The results of Ref. \cite{Motrunich2004} imply that the monopoles in the resulting $\U$ QSL have translation actions belonging to the nontrivial class of $\H^2_{\rho}[\Z^3; \U]$ with $\rho=-1$. Furthermore, this monopole should not carry spin-1/2 under SO(3) due to the absence of a nontrivial surface state of these fermions. Therefore, the gauged state is precisely $(\E_{f\half}\M_{b\trn})_-$.

\end{document}